\documentclass[aps,twocolumn,showpacs,superscriptaddress]{revtex4-1}
\usepackage{graphicx}
\usepackage{amsmath}
\usepackage{amssymb}
\usepackage{bm}
\usepackage{hyperref}
\usepackage{multirow}
\usepackage{appendix}
\usepackage{color}
\usepackage{placeins}

\newcommand{\nn}{\nonumber}

\begin{document}
\title{The Majorana spin in magnetic atomic chain systems}
\author{Jian Li}
\affiliation{
Department of Physics,
Princeton University,
Princeton, NJ 08544, USA
            }
\affiliation{
Institute for Natural Sciences, Westlake Institute for Advanced Study,
Hangzhou, Zhejiang Province, China
            }
\affiliation{
Westlake University,
Hangzhou, Zhejiang Province, China
            }
 \author{Sangjun Jeon}
\affiliation{
Department of Physics,
Princeton University,
Princeton, NJ 08544, USA
            }
 \author{Yonglong Xie}
\affiliation{
Department of Physics,
Princeton University,
Princeton, NJ 08544, USA
            }

\author{Ali Yazdani}
\affiliation{
Department of Physics,
Princeton University,
Princeton, NJ 08544, USA
            }

\author{B. Andrei Bernevig}
\affiliation{
Department of Physics,
Princeton University,
Princeton, NJ 08544, USA
            }
\affiliation{
Donostia International Physics Center, P. Manuel de Lardizabal 4, 20018 Donostia-San Sebasti\'{a}n, Spain
}
\affiliation{
Laboratoire Pierre Aigrain, Ecole Normale Sup\'{e}rieure-PSL Research University,
CNRS, Universit\'{e} Pierre et Marie Curie-Sorbonne Universit\'{e}s,
Universit\'{e} Paris Diderot-Sorbonne Paris Cit\'{e}, 24 rue Lhomond, 75231 Paris Cedex 05, France
}
\affiliation{
Sorbonne Universit\'{e}s, UPMC Univ Paris 06, UMR 7589, LPTHE, F-75005, Paris, France
}
\date{\today}
\begin{abstract}
In this paper, we establish that Majorana zero modes emerging from a topological band structure of a chain of magnetic atoms embedded in a superconductor can be distinguished from trivial localized zero energy states that may accidentally form in this system using spin resolved measurements. To demonstrate this key Majorana diagnostics, we study the spin composition of magnetic impurity induced in-gap Shiba states in a superconductor using a quantum impurity model (at the mean-field level). By examining the spin and spectral densities in the context of the Bogoliubov-de Gennes (BdG) particle-hole symmetry, we derive a sum rule that relates the spin densities of localized Shiba states with those in the normal state without superconductivity. Extending our investigations to ferromagnetic chain of magnetic impurities, we identify key features of the spin properties of the extended Shiba state bands, as well as those associated with a localized Majorana end mode when the effect of spin-orbit interaction is included. We then formulate a phenomenological theory for the measurement of the local spin densities with spin-polarized scanning tunneling microscopy (STM) techniques. By combining the calculated spin densities and the measurement theory, we show that spin-polarized STM measurements can reveal a sharp contrast in spin polarization between an accidentally-zero-energy trivial Shiba state and a Majorana zero mode in a topological superconducting phase in atomic chains. We further confirm our results with numerical simulations that address generic parameter settings. 
\end{abstract}
\maketitle

\section{Introduction}

Experimental breakthroughs \cite{mourik_signatures_2012, das_zero-bias_2012, rokhinson_fractional_2012, deng_anomalous_2012, nadj-perge_observation_2014, sun_majorana_2016} have advanced the research on Majorana zero modes (MZMs) \cite{wilczek_majorana_2009, alicea_new_2012, beenakker_search_2013, elliott_colloquium:_2015, beenakker_road_2016, lutchyn_realizing_2017} from appealing theoretical ideas \cite{kitaev_unpaired_2001, fu_superconducting_2008, nilsson_splitting_2008, lutchyn_majorana_2010, oreg_helical_2010, choy_majorana_2011, martin_majorana_2012, nadj-perge_proposal_2013, klinovaja_topological_2013, braunecker_interplay_2013, vazifeh_self-organized_2013, pientka_topological_2013, li_topological_2014} to an exciting new stage. Intensive efforts are being made in laboratories \cite{ruby_end_2015, qu_electric_2015, albrecht_exponential_2016, deng_majorana_2016, pawlak_probing_2016, feldman_high-resolution_2017, ruby_exploring_2017} towards realizing the full potential of MZMs, for example, to demonstrate non-abelian braiding statistics \cite{read_paired_2000, ivanov_non-abelian_2001, alicea_non-abelian_2011, aasen_milestones_2016, li_manipulating_2016}, and to ultimately perform topological quantum computation \cite{kitaev_fault-tolerant_2003, sarma_majorana_2015, nayak_non-abelian_2008}. Experimental evidence of MZMs is not only necessary to consolidate existing observations but also valuable for a deeper understanding of the current platforms. In the semiconductor nanowire Majorana platform \cite{lutchyn_majorana_2010, oreg_helical_2010, mourik_signatures_2012, das_zero-bias_2012, rokhinson_fractional_2012, deng_anomalous_2012}, for instance, observations of the exponential decay of MZMs, albeit by indirect means, have been recently reported \cite{albrecht_exponential_2016}. In the atomic chain Majorana platform \cite{nadj-perge_observation_2014, ruby_end_2015, choy_majorana_2011, martin_majorana_2012, nadj-perge_proposal_2013, klinovaja_topological_2013, braunecker_interplay_2013, vazifeh_self-organized_2013, pientka_topological_2013, li_topological_2014, kirsanskas_yu-shiba-rusinov_2015, schecter_self-organized_2016, christensen_spiral_2016, andolina_topological_2017} a series of recent high resolution measurements have placed more stringent bounds on the MZM splitting at low temperatures, revealed new spatial structures of MZMs, and established their equal electron-hole weights using spectroscopy with a superconducting scanning tunneling microscopy (STM) tip \cite{feldman_high-resolution_2017}. In this latter platform, a strong localization of the MZMs has been observed \cite{nadj-perge_observation_2014,  ruby_end_2015} with spatially resolved STM spectroscopy measurements, and is now theoretically well understood \cite{peng_strong_2015, li_topological_2014}. In all these Majorana platforms, the however small possibility that trivial end states can be accidentally tuned to zero energy and hence incorrectly identified as MZMs still exists \cite{liu_andreev_2017, danon_conductance_2017}. For example, in the atomic chain platform, it is possible that conventional localized Shiba states \cite{yu_bound_1965, shiba_classical_1968, rusinov_theory_1969} are accidentally tuned to nearly zero energy by a local potential at the end of a magnetic chain. Such a possibility, however improbable, cannot be excluded by energy resolved spectroscopic measurements alone but requires other types of diagnostics \cite{rosdahl_andreev_2017}. In this paper, we show that spin polarization \cite{sticlet_spin_2012, bjornson_spin-polarized_2015} can distinguish MZMs and trivial Shiba states and show how this is revealed by spin-polarized STM measurements \cite{wiesendanger_spin_2009}. Experimental demonstration that establishes this distinction has recently been accomplished for magnetic chains of Fe atoms on the surface of Pb \cite{Jeong_2017_to_appear}.

The paper is organized as follows. In Section \ref{sec:toy}, we illustrate the key idea of this paper, using a one-body toy model for one proximitized magnetic impurity. In Section \ref{sec:single}, we examine the spin properties of Shiba states induced by a single magnetic impurity using a fully quantum mechanical model of such a impurity hybridized with a generic conventional s-wave superconductor. A key result discussed in this section is the relation between the spin densities induced locally by the impurity in the normal state and those associated with the in-gap Shiba states in the superconducting state. These results show how such localized states differ in their spin properties from a MZM that emerges in the topological phase of the atomic chains, and set the stage to understand the spin properties of Shiba bands induced by a ferromagnetic chain of magnetic atoms in a superconductor discussed in Section \ref{sec:chain}. In Section \ref{sec:chain}, we derive in detail the spin properties of a MZM when spin-orbit coupling drives the system into a topological superconducting phase. To relate our theory to a recent experiment \cite{Jeong_2017_to_appear}, we develop a phenomenological theory in Section \ref{sec:stm} that captures how spin-polarized STM probes the properties of the in-gap states in a superconductor. An important result established in this section is a diagnostic test: unlike a MZM, a trivial zero mode with wavefunction support on the end of the chain would show no contrast in spin-polarized STM measurements performed under typical constant current conditions. In Section \ref{sec:sim}, we complement our analytical results with numerical simulations, using generic but realistic parameters. Aside from confirming our analytical results, our simulations allow us to demonstrate the utility of our Majorana spin diagnostic test.

\section{Toy model}\label{sec:toy}

\begin{figure}[h]
  \centering
  \includegraphics[width=0.4\textwidth]{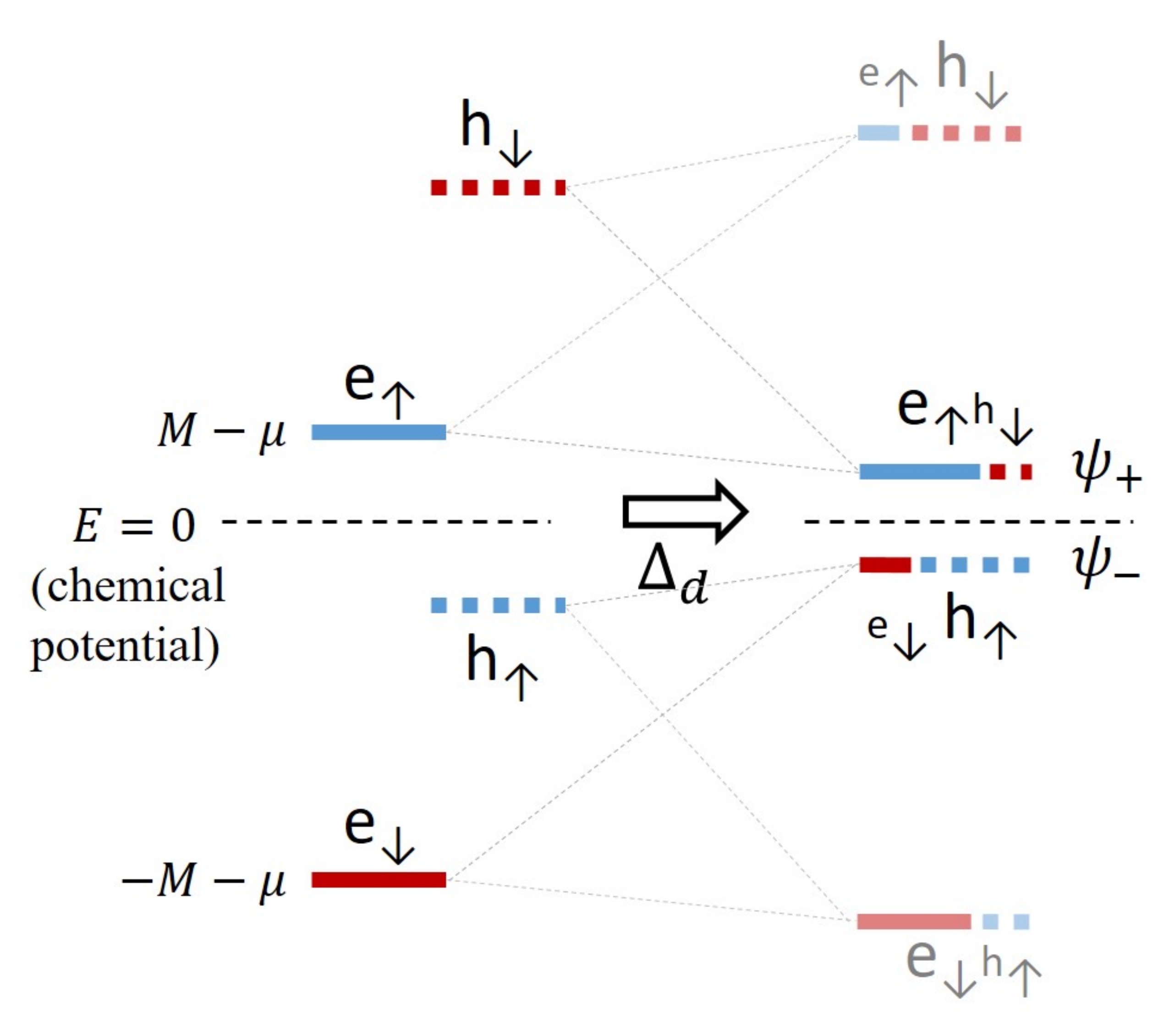}
  \caption{Illustration of the toy model. The solid-line segments represent the electron components corresponding to the $d$ operators, and the broken-line segments represent the hole components corresponding to the $\bar{d}$ operators. The quasiparticle states in the presence of spin-singlet pairing are superpositions of electron and hole components with opposite spins. The combination of an exchange field and a spin-singlet pairing generically results in quasiparticle states (e.g. $\psi_{+}$ and $\psi_{-}$ on the right; see text for the details of the states) of opposite electronic spin polarization and asymmetric spectral weights at opposite energies.}
  \label{fig:toy}
\end{figure}

In order to demonstrate the main idea of this paper, we start with a toy model comprising only a single site and two spins. The Hamiltonian is given by
\begin{align}
  &\hat{H} =
  \begin{pmatrix}
    \bm{d}^\dagger & \bar{\bm{d}}^\dagger
  \end{pmatrix}
  (H_d - \Delta_d\sigma_0\otimes\tau_x)
  \begin{pmatrix}
    \bm{d} \\ \bar{\bm{d}}
  \end{pmatrix}, \label{eq:hamtoy} \\
  & H_d = M\sigma_z\otimes\tau_0-\mu\sigma_0\otimes\tau_z, \label{eq:Hd0}
\end{align}
where $M$, $\mu$ and $\Delta_d$ are real, non-negative parameters representing the (magnetization) exchange energy, the chemical potential and the (induced) pairing potential, respectively; $\bm{d}^\dag = (d_{\uparrow}^\dag, d_{\downarrow}^\dag)$ and $\bar{\bm{d}}^\dag = (d_{\downarrow}, -d_{\uparrow})$ represent the Nambu particle (without bar) and hole (with bar) creation operators,  respectively; $\sigma_i$ ($i=0,x,y,z$) stands for the Pauli matrices for spin, and $\tau_i$ ($i=0,x,y,z$) stands for the Pauli matrices for the Nambu particle-hole spinors. From now on we will drop the identity matrices $\sigma_0$ and $\tau_0$ where no ambiguity will arise.

The above Hamiltonian is easily solved and the two low energy eigenstates are given by (see Fig.~\ref{fig:toy})
\begin{align}
  &E_{\pm} = \pm (M - \sqrt{\mu^2+\Delta_d^2}), \label{eq:E_toy}\\
  &\psi_{+} =
  \begin{pmatrix}
    \cos\frac{\theta}{2} \\ 0 \\ \sin\frac{\theta}{2} \\ 0
  \end{pmatrix},\;
  \psi_{-} =
  \begin{pmatrix}
    0 \\ \sin\frac{\theta}{2} \\ 0 \\ -\cos\frac{\theta}{2}
  \end{pmatrix}, \label{eq:psi_toy}
\end{align}
where, by definition, $\cos\theta={\mu}/{\sqrt{\mu^2+\Delta_d^2}}$ and $\sin\theta={\Delta_d}/{\sqrt{\mu^2+\Delta_d^2}}$. When both low energy states are exactly at zero energy, which requires $M = \sqrt{\mu^2+\Delta_d^2}$, we can artificially construct Majorana states by superposing $\psi_{+}$ and $\psi_{-}$ as (in the $\bm{d}$, $\bar{\bm{d}}$ basis)
\begin{subequations}\label{eq:majorana_toy}
\begin{align}
  \chi_1 &= \frac{1}{\sqrt{2}}(e^{-i\varphi}\psi_+ + e^{i\varphi}\psi_-), \\
  \chi_2 &= \frac{i}{\sqrt{2}}(e^{-i\varphi}\psi_+ - e^{i\varphi}\psi_-),
\end{align}
\end{subequations}
where $\varphi$ stands for the gauge freedom in constructing the Majorana states.

In this paper, we are particularly interested in the particle components of the Nambu spinors because of their relevance to the STM measurement in the single-electron sequential tunneling regime. To this end we define spin densities for an eigenstate $\psi$ to be
\begin{align}
  \rho_{\uparrow/\downarrow}(\psi) \equiv \langle\psi | p_{\uparrow/\downarrow} | \psi\rangle, \label{eq:rho_def_eig}
\end{align}
where
\begin{align}
  p_{\uparrow/\downarrow}=\frac{1}{2}(\sigma_0\pm\sigma_z)\otimes\frac{1}{2}(\tau_0+\tau_z) \label{eq:p_def}
\end{align}
is the projector to the spin-$\uparrow$/$\downarrow$ particle component, respectively. From Eq.~\eqref{eq:psi_toy}, it is clear that $\psi_{+}$ ($\psi_{-}$) only contains nonvanishing spin-$\uparrow$ (spin-$\downarrow$) component:
\begin{subequations}\label{eq:rho_toy}
\begin{align}
 & \rho_{\uparrow}(\psi_+) = \frac{1}{2}\left(1+\frac{\mu}{\sqrt{\mu^2+\Delta_d^2}}\right),\quad \rho_{\downarrow}(\psi_+) = 0;\\
 & \rho_{\downarrow}(\psi_-) = \frac{1}{2}\left(1-\frac{\mu}{\sqrt{\mu^2+\Delta_d^2}}\right),\quad \rho_{\uparrow}(\psi_-)=0\,. 
\end{align}
\end{subequations}
In other words, the two low energy states $\psi_{\pm}$ are both fully spin-polarized with opposite spin polarization (see Fig.~\ref{fig:toy}). Therefore we may formally write down $\rho_{\uparrow}$ ($\rho_{\downarrow}$) as a function of energy $E_{+}$ ($E_{-}$) that is associated with $\psi_{+}$ ($\psi_{-}$), hence we have
\begin{subequations}\label{eq:rhoE_toy}
\begin{align}
 & \rho_{\uparrow}(E) \equiv \rho_{\uparrow}(\psi_+|_{E_+=E}) = \frac{1}{2}+\frac{1}{2}\sqrt{1-\frac{\Delta_d^2}{(M-E)^2}}\,, \\
 & \rho_{\downarrow}(E) \equiv \rho_{\downarrow}(\psi_-|_{E_-=E}) = \frac{1}{2}-\frac{1}{2}\sqrt{1-\frac{\Delta_d^2}{(M+E)^2}}\,.
\end{align}
\end{subequations}
By assuming $\Delta_d\ll M\pm E$, which is always true if $\max(\Delta_d, |\mu-M|)\ll M$, we obtain $\rho_{\uparrow}(E) \simeq 1-{\Delta_d^2}/4(M-E)^2$ and $\rho_{\downarrow}(E) \simeq {\Delta_d^2}/4(M+E)^2$. This implies that the spin densities are dominated by the spin polarization of the original state that is closer to the chemical potential. In the case of Fig.~\ref{fig:toy}, this is the electronic spin-$\uparrow$ state, whether occupied or unoccupied.

Moreover, the difference between the two spin densities satisfies
\begin{align}
  \delta\rho(E) &\equiv \rho_{\uparrow}(E) - \rho_{\downarrow}(E) \\
  &= \frac{1}{2}\left[\sqrt{1-\frac{\Delta_d^2}{(M-E)^2}}+\sqrt{1-\frac{\Delta_d^2}{(M+E)^2}}\right] \\
  &\le \sqrt{1-\frac{1}{2}\left[\frac{\Delta_d^2}{(M-E)^2}+\frac{\Delta_d^2}{(M+E)^2}\right]} \\
  &\le \sqrt{1-\frac{\Delta_d^2}{M^2}} = \delta\rho(E=0)\,.
\end{align}
Namely, $\delta\rho(E)$ reaches its maximum at $E=0$. On the other hand, for the artificial Majorana states $\chi_{1,2}$ in Eq.~\eqref{eq:majorana_toy}, we straightforwardly find
\begin{align}
  \delta\rho(\chi_{1,2}) &\equiv \rho_{\uparrow}(\chi_{1,2})-\rho_{\downarrow}(\chi_{1,2}) \\
  &= \frac{1}{2}\sqrt{1-\frac{\Delta_d^2}{M^2}} = \frac{1}{2}\delta\rho(E=0)\,.
\end{align}
Apart from a factor of $1/2$ owing to the fact that the spin densities at $E=0$ are split equally to two Majorana states, we see that $\delta\rho$ for the Majorana states corresponds to the maximum of $\delta\rho(E)$, at $E=0$, associated with generic eigenstates $\psi_{\pm}$. This behavior, as we see in later sections, where Shiba and Majorana states are considered, is key for a MZM to be distinguished from trivial Shiba states in spin-polarized STM measurement.

Incidentally, if we choose a different orientation for the spin basis, which, for example, amounts to replacing $\sigma_z$ in Eq.~\eqref{eq:p_def} by $\sigma_x$ or $\sigma_y$ and denoting the corresponding $\delta\rho\equiv\rho_{\uparrow}-\rho_{\downarrow}$ by $\delta\rho_x$ or $\delta\rho_y$, it is easy to verify that $\delta\rho_{x/y}(\psi)=0$ if $\psi = \psi_{\pm}$, whereas $\delta\rho_x(\chi_1) = -\delta\rho_x(\chi_2) = \frac{\Delta_d}{2M}\cos2\varphi$, and $\delta\rho_y(\chi_1) = -\delta\rho_y(\chi_2) = \frac{\Delta_d}{2M}\sin2\varphi$. This indicates that Majorana states, albeit artificial in the toy model, may acquire finite in-plane spin polarization in contrast to trivial quasiparticle states represented by $\psi_{\pm}$ \cite{sticlet_spin_2012}. Such in-plane spin polarization, however, is very weak by realistic measure as $\Delta_d\ll M$; therefore we will focus on only the spin polarization along the magnetization ($z$) in the following of this paper.

\section{Single magnetic impurity model}\label{sec:single}

We now build up towards our Shiba chain model by considering the model that consists of a 2D or 3D bulk superconductor coupled to a single quantum magnetic impurity. We assume the superconductor to be infinite (without any surface) in all its dimensions and neglect spin-orbit coupling for simplicity. This hybrid system can be described by a Bogoliubov-de Gennes (BdG) Hamiltonian
\begin{align}
  &\hat{H} = \hat{H}_s + \hat{H}_d + \hat{H}_T, \label{eq:ham_full0}\\
  &\hat{H}_s = \int{d\bm{k}}\,
  \begin{pmatrix}
    \bm{c}_{\bm{k}}^\dagger & \bar{\bm{c}}_{\bm{k}}^\dagger
  \end{pmatrix} H_s(\bm{k})
  \begin{pmatrix}
    \bm{c}_{\bm{k}} \\ \bar{\bm{c}}_{\bm{k}}
  \end{pmatrix}, \\
  &\hat{H}_d =
  \begin{pmatrix}
    \bm{d}^\dagger & \bar{\bm{d}}^\dagger
  \end{pmatrix}H_d
  \begin{pmatrix}
    \bm{d} \\ \bar{\bm{d}}
  \end{pmatrix}, \\
  &\hat{H}_T = \int{d\bm{r}}\,
  \begin{pmatrix}
    \bm{c}_{\bm{r}}^\dagger & \bar{\bm{c}}_{\bm{r}}^\dagger
  \end{pmatrix}
  V\delta(\bm{r})\tau_z
  \begin{pmatrix}
    \bm{d} \\ \bar{\bm{d}}
  \end{pmatrix} + h.c., \label{eq:ham_coup0}
\end{align}
with
\begin{align}
  & H_s(\bm{k}) = (t_s k^2-\mu_s)\tau_z+\Delta\tau_x. \label{eq:Hs}
\end{align}
Here, $t_s$, $\mu_s$ and $\Delta$ stand for the band-width parameter, the chemical potential and the (real) pairing potential for the superconductor, respectively; $\bm{c}_{\bm{r}}^\dag = (c_{\bm{r}\uparrow}^\dag, c_{\bm{r}\downarrow}^\dag)$ and $\bar{\bm{c}}_{\bm{r}}^\dag = (c_{\bm{r}\downarrow}, -c_{\bm{r}\uparrow})$ represent the Nambu particle and hole creation operators for the superconductor,  respectively; $\bm{c}_{\bm{k}}=\int_{-\infty}^{\infty}{d\bm{r}} e^{-i\bm{k}\cdot\bm{r}} \bm{c}_{\bm{r}}$ and $\bar{\bm{c}}_{\bm{k}}=\int_{-\infty}^{\infty}{d\bm{r}} e^{-i\bm{k}\cdot\bm{r}} \bar{\bm{c}}_{\bm{r}}$ are the Fourier transforms of $\bm{c}_{\bm{r}}$ and $\bar{\bm{c}}_{\bm{r}}$, respectively; $H_d$, as well as $\bm{d}$ and $\bar{\bm{d}}$, is similarly defined as in Eqs.~\eqref{eq:hamtoy} and \eqref{eq:Hd0}. With the tunneling Hamiltonian \eqref{eq:ham_coup0}, we have assumed the single magnetic impurity to be sitting at $\bm{r}=0$.

Throughout this paper we will focus on the spin densities on the $d$-orbitals -- namely, the spin densities evaluated with $\bm{d}$ and $\bm{d}^\dagger$ operators, $s_i = \langle\psi|\bm{d}^\dag \sigma_i \bm{d}|\psi\rangle \qquad (i=0,x,y,z)$. The reason for this is twofold: first, the experimental technique considered in this paper is STM, which measures locally with atomic resolution; second, although the major weight of a Shiba state is distributed in the superconductor, as we will see at the end of this section, the length scale of the distribution is given by the superconducting coherence length which is generally large ($\sim$ 80 nm in Pb, for example), such that the local weight of a Shiba state is small on a superconductor atom compared with that on the magnetic adatom (such as Fe) $d$-orbitals.

The retarded Green's function for the $d$-orbital degrees of freedom is given by (see Appendix \ref{app:GF})
\begin{align}
  &G_d(E^+) = \left[E^+ - H_d + v\frac{E^+\tau_0+\Delta\tau_x}{\sqrt{\Delta^2-(E^+)^2}}\right]^{-1}, \label{eq:Gd0}\\
  &v = \pi \rho_s V^2, \label{eq:v0}
\end{align}
where $E^\pm = E \pm i\eta$ with $\eta$ a positive infinitesimal, and $\rho_s$ is the normal DOS of the superconductor at its Fermi energy ($\rho_s = \sqrt{\mu_s/t_s}/4\pi^2 t_s$ if the superconductor is 3D, and $\rho_s = 1/4\pi t_s$ if the superconductor is 2D). Throughout this paper we will assume $M, v\gg\Delta$. Note that Eq.~\eqref{eq:Gd0} is valid for the full energy range (below or above the superconducting gap) as long as we adopt the convention of taking the square root such that $\text{Re}(\sqrt{\Delta^2-(E^+)^2})>0$. Peng et al. \cite{peng_strong_2015} have previously explored the limit of when $|E|\ll\Delta$ which leads to a deep-Shiba-limit effective Hamiltonian of the single impurity problem (see Appendix \ref{app:single}).

\begin{figure}
  \centering
  \includegraphics[width=0.5\textwidth]{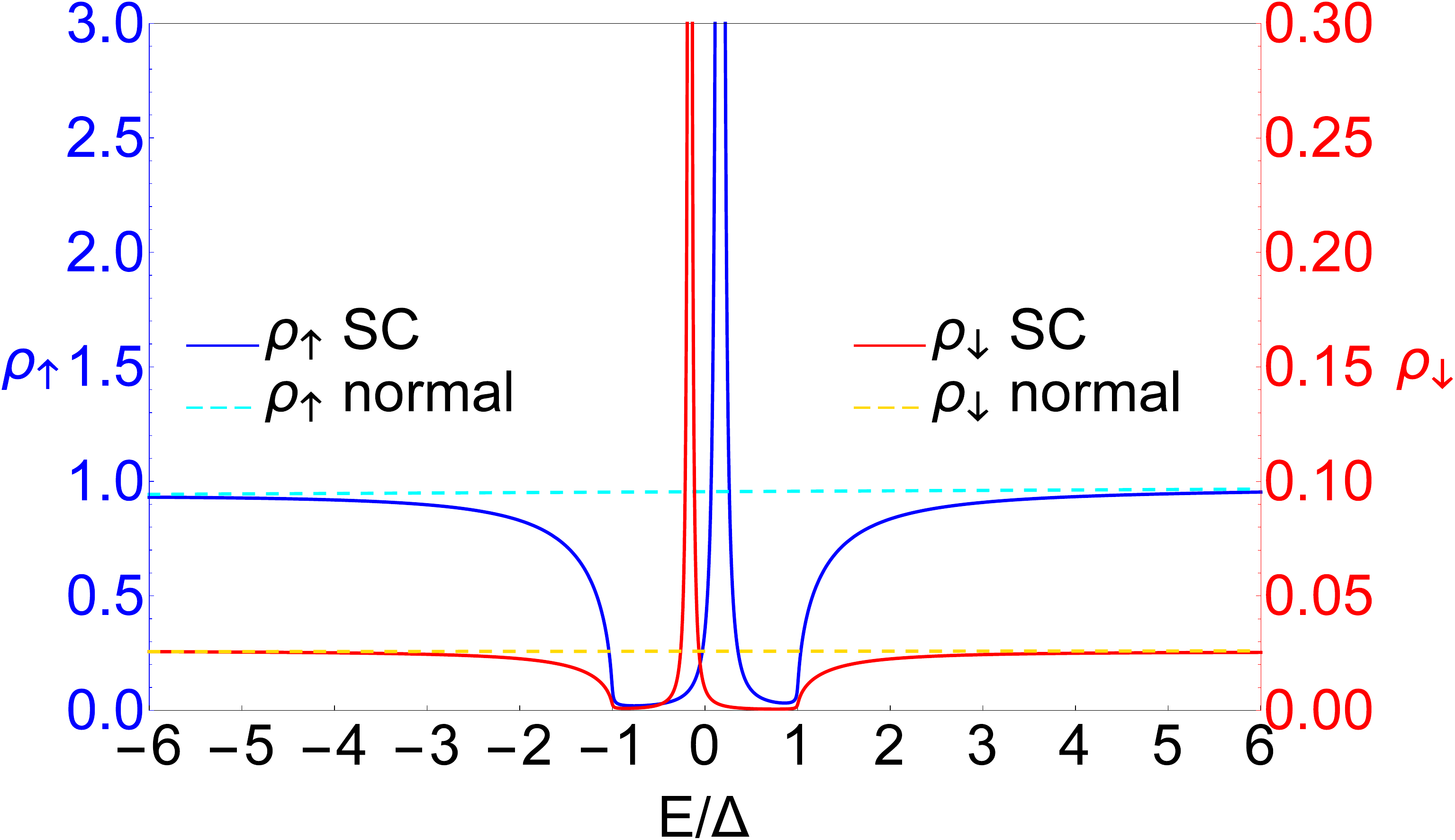}
  \caption{The spin densities as a function of energy in the superconducting state (solid lines) and in the normal state (broken lines). The sharp peaks inside the superconducting gap correspond to the Shiba states. The parameters used here are: $M=1$, $\mu=0.9$, $v=0.3$, $\Delta=0.001$ and $\eta=1\mathrm{e}-5$.}
  \label{fig:rhos_single}
\end{figure}

\begin{figure}
  \centering
  \includegraphics[width=0.45\textwidth]{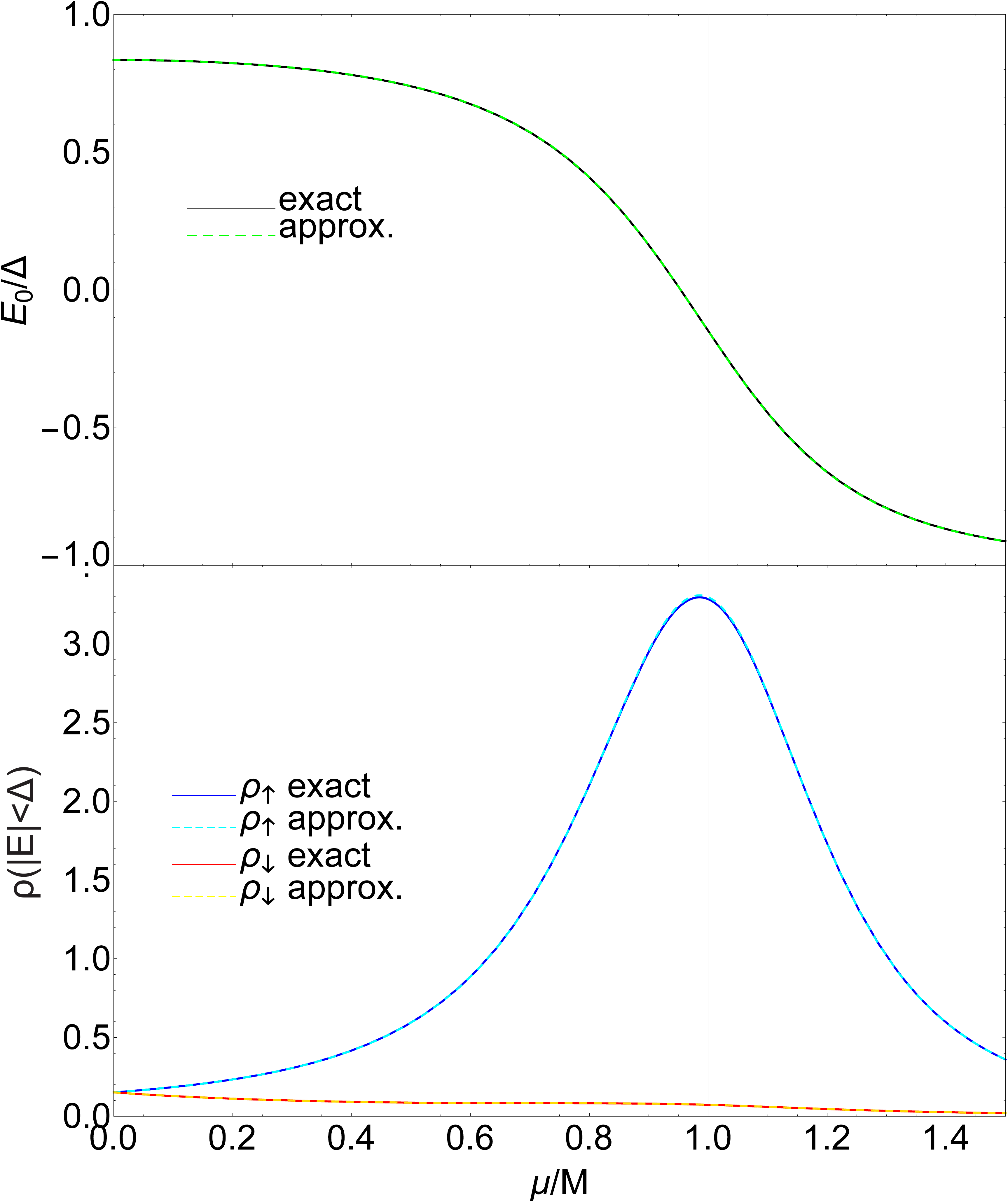}
  \caption{The energy $E_0$ (upper panel), and the spin densities (lower panel), of the Shiba states, as a function of the chemical potential $\mu$. The spin densities shown here are the integrated $\rho_{\uparrow/\downarrow}(E)$ over the subgap energy range. In both panels, the exact values are obtained by numerically solving Eq.~\eqref{eq:Gd0}; the approximate values are obtained from Eqs.~\eqref{eq:E0_shiba_main} and \eqref{eq:rho_subgap_main}, respectively. The parameters used here are: $M=1$, $v=0.3$, $\Delta=0.001$ and $\eta=1\mathrm{e}-5$.}
  \label{fig:E0_rhos}
\end{figure}

The $d$-orbital spin densities, defined as
\begin{align}
  \rho_{\uparrow/\downarrow}(E) = \text{Tr}[p_{\uparrow/\downarrow}A_d(E)], \label{eq:rho_def}
\end{align}
with $p_{\uparrow/\downarrow}$ given in Eq.~\eqref{eq:p_def} and $A_d(E) = \lim_{\eta\rightarrow0}\frac{i}{2\pi}[G_d(E^+) - G_d(E^-)]$, can be obtained from Eq.~\eqref{eq:Gd0} (see Appendix \ref{app:single}) to be:
\begin{align}
  &\rho_{\uparrow/\downarrow}(E) = \frac{|v_E| [(E \mp M - \mu)^2 + v^2] / \pi}{[(E \mp M)^2 - \mu^2 - v^2]^2 + 4 |v_E|^2 (E \mp M)^2}, \nn\\
  &\hspace{0.32\textwidth}\text{if } |E|\ge\Delta\,; \label{eq:rho_supergap_main}\\
  &\rho_{\uparrow/\downarrow}(E) \simeq \frac{v}{(\mu \mp M)^2 + v^2} \sqrt{\Delta^2 - E_0^2} \;\delta(E \mp E_0), \nn\\
  &\hspace{0.32\textwidth}\text{if } |E|<\Delta\,. \label{eq:rho_subgap_main}
\end{align}
where
\begin{align}
  &v_E = vE/\sqrt{\Delta^2-E^2}, \label{eq:vE}\\
  &E_0 \simeq \Delta\,\frac{M^2-\mu^2-v^2}{\sqrt{(M^2-\mu^2-v^2)^2+4M^2v^2}}, \label{eq:E0_shiba_main}
\end{align}
In Fig.~\ref{fig:rhos_single}, we show the spin densities obtained directly from Eq.~\eqref{eq:rho_def} with a small finite $\eta$, which accounts for the finite width of the delta functions; in the upper panel of Fig.~\ref{fig:E0_rhos}, we show the energy of a Shiba state, $E_0$, as a function of $\mu$, from the exact solution of the poles in Eq.~\eqref{eq:Gd0}, as well as from the approximate expression Eq.~\eqref{eq:E0_shiba_main} in the limit $\Delta\ll M$.

Eq.~\eqref{eq:rho_supergap_main} implies that (see Appendix \ref{app:single}), in the supergap regime, $\rho_{\uparrow/\downarrow}(E)$ increase as $\sqrt{|E|/\Delta-1}$ from the gap edge and converge to their normal-state values at an energy far from the gap:
\begin{align}
   &\rho_{\uparrow/\downarrow}(E) \simeq \rho_{\uparrow/\downarrow}^{(N)}(E) \hspace{0.05\textwidth}\text{if } |E|\gg\Delta, \label{eq:rho_app_normal_main}
\end{align}
where
\begin{align}
   &\rho_{\uparrow/\downarrow}^{(N)}(E) \equiv \frac{v / \pi}{(E \mp M + \mu)^2 + v^2}\label{eq:rho_normal_main}
\end{align}
are the normal-state spin densities in the full energy range ($\Delta$ is irrelevant in the normal state). This convergence is certainly expected and can be seen straightforwardly from Eq.~\eqref{eq:Gd0} by noticing that $|E|\gg\Delta$ implies $G_d(E^+) \simeq (E - H_d + iv)^{-1}$ with vanishing pairing terms. In Fig.~\ref{fig:rhos_single}, we plot in broken lines the normal-state spin densities $\rho_{\uparrow/\downarrow}^{(N)}$ in the same energy range as the superconducting-state spin densities, where the convergence is clearly seen. The energy range shown in Fig.~\ref{fig:rhos_single} is small compared with $|M\pm\mu|$ or $v$; therefore $\rho_{\uparrow/\downarrow}^{(N)}(E)$ are roughly constants. The superconducting-state spin densities $\rho_{\uparrow/\downarrow}(E)$ approach these constants under the condition $\Delta \ll |E| \ll \max(|M-\mu|,v)$.

In the more interesting subgap regime, Eq.~\eqref{eq:rho_subgap_main} shows that the Shiba state occurring at $E = +E_0\;(-E_0)$ has only nonvanishing spin-$\uparrow$($\downarrow$) electronic components. More importantly, the total spin densities inside the gap are given by (see also the lower panel of Fig.~\ref{fig:E0_rhos})
\begin{align}
  &\int_{-\Delta}^{\Delta}\rho_{\uparrow/\downarrow}(E)\, dE \simeq \pi\sqrt{\Delta^2 - E_0^2}\; \rho_{\uparrow/\downarrow}^{(N)}, \label{eq:sum_rule0} \\
  &\rho_{\uparrow/\downarrow}^{(N)} \equiv \rho_{\uparrow/\downarrow}^{(N)}(E=0),
\end{align}
where the second equation stands for a shorthand notation and $\rho_{\uparrow/\downarrow}^{(N)}(E)$ are defined in Eq.~\eqref{eq:rho_normal_main}. This represents a sum rule regarding the redistribution of $d$-orbital spin densities, as well as its spectral density, upon the opening of a superconducting gap in the host. Moreover, in a larger energy range $|E|\le E_c$ where $\Delta\ll E_c\ll v,M$, the sum rule resumes a more transparent form (see Appendix \ref{app:single})
\begin{align}
  \int_{-E_c}^{E_c}\rho_{\uparrow/\downarrow}(E)\, dE \simeq 2E_c\,\rho_{\uparrow/\downarrow}^{(N)}. \label{eq:sum_rule_full}
\end{align}
Note that the ratio of the integrated $\rho_{\uparrow}$ to the integrated $\rho_{\downarrow}$, either in Eq.~\eqref{eq:sum_rule0} or in Eq.~\eqref{eq:sum_rule_full}, is always equal to the ratio of the normal-state spin densities at the chemical potential.

A particularly important limit in the subgap regime is when $|E|\ll\Delta$. In this limit, as has been point out by Peng et al. \cite{peng_strong_2015}, Eq.~\eqref{eq:Gd0} becomes (to linear order in $E^+/\Delta$)
\begin{align}
  &G_d(E^+) \simeq \left[E^+(1+v/\Delta) - H_d + v\tau_x\right]^{-1}, \nn\\
  &\hspace{0.32\textwidth}\text{if } |E|\ll\Delta,\, \label{eq:Gd0_Esmall}
\end{align}
which leads to a deep-Shiba-limit effective Hamiltonian
\begin{align}
  \tilde{H}_{d} = \frac{\Delta}{\Delta+v}\,(H_d - v\tau_x). \label{eq:ham_deep_shiba}
\end{align}
This effective Hamiltonian has the same form of the toy model Eq.~\eqref{eq:hamtoy}, except that $v$ has replaced $\Delta_d$, and there is an additional scaling factor $\Delta/(\Delta+v)$ which represents the portion of the weight of a Shiba state actually on the $d$-orbitals \cite{peng_strong_2015}. Here, we emphasize that for self-consistency the low-energy eigenstates of the Hamiltonian~\eqref{eq:ham_deep_shiba} correspond to Shiba states only if the eigenvalues associated with these eigenstates are small compared to $\Delta$, which amounts to the condition $|M-\sqrt{\mu^2+v^2}| \ll v$. Under such a condition, and to the leading order in $(M-\sqrt{\mu^2+v^2})/v$, it is straightforward to show that Eqs.~\eqref{eq:E0_shiba_main} and \eqref{eq:rho_subgap_main} reduce to Eqs.~\eqref{eq:E_toy} and \eqref{eq:rho_toy}, respectively (see Appendix \ref{app:single}).

Given that Shiba states are the only way to produce a localized in-gap state in a conventional superconductor, the calculations of this section will provide an important spin signature of such states if they were to form accidentally at zero energy. As we describe below, the analytical results of this section can be combined with an STM measurement theory to provide a key difference between the measured spin contrast from such states and those from a MZM that emerge in a topological magnetic chain.

\section{Magnetic impurity chain model}\label{sec:chain}

Now we are in the position to discuss a 3D model composed of a bulk superconductor and a magnetic impurity chain:
\begin{align}
  &\hat{H} = \hat{H}_s + \hat{H}_d + \hat{H}_T, \label{eq:ham_full}\\
  &\hat{H}_s = \int{d\bm{k}}\,
  \begin{pmatrix}
    \bm{c}_{\bm{k}}^\dagger & \bar{\bm{c}}_{\bm{k}}^\dagger
  \end{pmatrix} H_s(\bm{k})
  \begin{pmatrix}
    \bm{c}_{\bm{k}} \\ \bar{\bm{c}}_{\bm{k}}
  \end{pmatrix}, \\
  &\hat{H}_d = \int{d k_x}\,
  \begin{pmatrix}
    \bm{d}_{k_x}^\dagger & \bar{\bm{d}}_{k_x}^\dagger
  \end{pmatrix}H_d(k_x)
  \begin{pmatrix}
    \bm{d}_{k_x} \\ \bar{\bm{d}}_{k_x}
  \end{pmatrix}, \label{eq:ham_Hd} \\
  &\hat{H}_T = \int{d k_x}\,{dy}\,{dz}\,
  \begin{pmatrix}
    \bm{c}_{k_x,y,z}^\dagger & \bar{\bm{c}}_{k_x,y,z}^\dagger
  \end{pmatrix}
  V\delta(y)\delta(z)\tau_z
  \begin{pmatrix}
    \bm{d}_{k_x} \\ \bar{\bm{d}}_{k_x}
  \end{pmatrix} \nn\\
  &\hspace{0.35\textwidth}+ h.c., \label{eq:ham_coup}
\end{align}
where $H_s(\bm{k})$ is given by Eq.~\eqref{eq:Hs}, and
\begin{align}
  H_d(k_x) = M\sigma_z + [\xi_d(k_x) - \mu +\xi_{SO}(k_x)\sigma_y]\otimes\tau_z, \label{eq:Hd}
\end{align}
with $\xi_d$ a real symmetric function of $k_x$ and $\xi_{SO}$ a real anti-symmetric function of $k_x$, representing the spin-independent and the spin-orbit-coupling energies, respectively; $\bm{d}_{x}^\dag = (d_{x\uparrow}^\dag, d_{x\downarrow}^\dag)$, $\bar{\bm{d}}_{x}^\dag = (d_{x\downarrow}, -d_{x\uparrow})$, $\bm{d}_{k_x}=\int_{-\infty}^{\infty}{dx}\, e^{-ik_x x} \bm{d}_{x}$ and $\bar{\bm{d}}_{k_x}=\int_{-\infty}^{\infty}{dx}\, e^{-ik_x x} \bar{\bm{d}}_{x}$. For simplicity, we have assumed that: first, the chain is embedded in a 3D bulk superconductor with no surface; second, spin-orbit coupling, although mainly induced from the host material in reality, is added only to the Hamiltonian for the chain. Since $k_x$ is a good quantum number in the above Hamiltonian, we can solve the model for each fixed $k_x$ separately such that the problem is reduced to 2D with a single magnetic impurity, as we have solved in the previous section. With this dimensional reduction, we define an effective $k_x$-dependent chemical potential $\mu_d(k_x) = \mu - \xi_d(k_x)$, and modify Eq.~\eqref{eq:v0} to be $v(k_x) = \pi\rho_s(k_x)V^2$ with $\rho_s(k_x)$ the normal DOS of the substrate at its Fermi energy with a fixed $k_x$.

Before we proceed, let us first examine the behavior of $v(k_x)$. By assuming the normal-state Hamiltonian of the host to be $H_s^{(N)}(\bm{k}) = t_s k^2 - \mu_s$ [cf. Eq.~\eqref{eq:Hs}], we obtain $\rho_s(k_x)$ to be a rectangular function
\begin{align}
  \rho_s(k_x)=\left\{
  \begin{array}{lr}
    1/4\pi t_s, & \text{if } |k_x|\le k_c\,; \\
    0, & \text{if } |k_x|>k_c\,,
  \end{array}
  \right.
\end{align}
where $k_c = \sqrt{\mu_s/t_s}$ is the Fermi wave-vector for the bulk superconductor. Therefore $v(k_x)$ also appears to be a rectangular function that is given by a constant $v=V^2/4t_s$ inside the cutoff momentum range $[-k_c, k_c]$ and 0 otherwise. In the rest of this paper, we shall assume the range $|k_x|\le k_c$ to be sufficiently large such that the low energy (smaller than or comparable to $\Delta$) states of the pristine $d$-orbital bands always fall into this momentum range -- outside this momentum range, because of the vanishing $\rho_s(k_x)$ and hence the vanishing $v(k_x)$, the self-energy term in $G_d(E^+)$ in Eq.~\eqref{eq:Gd0} is also vanishing, thus the $d$-orbital states outside $[-k_c, k_c]$ become irrelevant at low energy by our assumption.

\subsection{Shiba bands in the absence of spin-orbit coupling}\label{ssec:shiba_band}

We start with the limit of vanishing spin-orbit coupling, namely, $\forall k_x: \xi_{SO}(k_x)=0$. The Shiba band dispersion relation is given by replacing $\mu$ with $\mu_d(k_x)$ in Eq.~\eqref{eq:E0_shiba_main}:
\begin{align}
  &E_0(k_x) \simeq \Delta\,\frac{M^2-\mu_d(k_x)^2-v^2}{\sqrt{(M^2-\mu_d(k_x)^2-v^2)^2+4M^2v^2}}. \label{eq:E0_shiba_band}
\end{align}
Similarly the $k_x$-dependent spin densities $\rho_{\uparrow/\downarrow}(k_x, E)$ can be obtained from Eqs.~\eqref{eq:rho_supergap_main} and \eqref{eq:rho_subgap_main}. The total spin densities are then given by
\begin{align}
  \rho_{\uparrow/\downarrow}(E) = \int_{-k_c}^{k_c} dk_x\, \rho_{\uparrow/\downarrow}(k_x, E). \label{eq:rho_E}
\end{align}
In the normal-state limit $|E|\gg\Delta$, from Eqs.~\eqref{eq:rho_app_normal_main} and \eqref{eq:rho_normal_main} we have
\begin{align} \label{eq:rho_band_normal}
  \rho_{\uparrow/\downarrow}(E) \simeq \int_{-k_c}^{k_c} dk_x\, \frac{v / \pi}{[E \mp M + \mu_d(k_x)]^2 + v^2}, 
\end{align}
which is an integration of normal-state spin densities over the momentum range $[-k_c,k_c]$ where the magnetic chain is strongly hybridized with the superconductor. In other words, the spin densities in the normal-state limit are contributed by an extended momentum range of the pristine $d$-orbital bands with broadening $v$. In the subgap regime $|E|<\Delta$, by using Eq.~\eqref{eq:rho_subgap_main} we have
\begin{align}
  &\rho_{\uparrow/\downarrow}(E) = \int_{-k_c}^{k_c} dk_x\, \frac{v\sqrt{\Delta^2 - E_0(k_x)^2}}{[\mu_d(k_x) \mp M]^2 + v^2} \;\delta[E \mp E_0(k_x)] \nn\\
  &\quad= \sum_{k_i^{\pm}} \frac{v\sqrt{\Delta^2 - E^2}}{[\mu_d(k_i^{\pm}) \mp M]^2 + v^2}\,\left|\frac{1}{\partial E_0/\partial k_x}\right|_{k_x=k_i^{\pm}}, \label{eq:rho_band_org}
\end{align}
where $k_i^{\pm}(E)$ are the solutions of the equations $E_0(k_x) = \pm E$, respectively for the two signs, in the range of $[-k_c, k_c]$. By definition, we have $k_i^+(+E) = k_i^-(-E)$ for all $E$, which is a manifestation of particle-hole symmetry. In contrast to the normal-state case, Eq.~\eqref{eq:rho_band_org} shows that the subgap spin densities at any specific energy ($|E|<\Delta$) are only contributed by a small set of momenta ($k_i^{\pm}(E)$) from the pristine $d$-orbital bands, and hence can vary strongly with energy.

\begin{figure}
  \centering
  \includegraphics[width=0.5\textwidth]{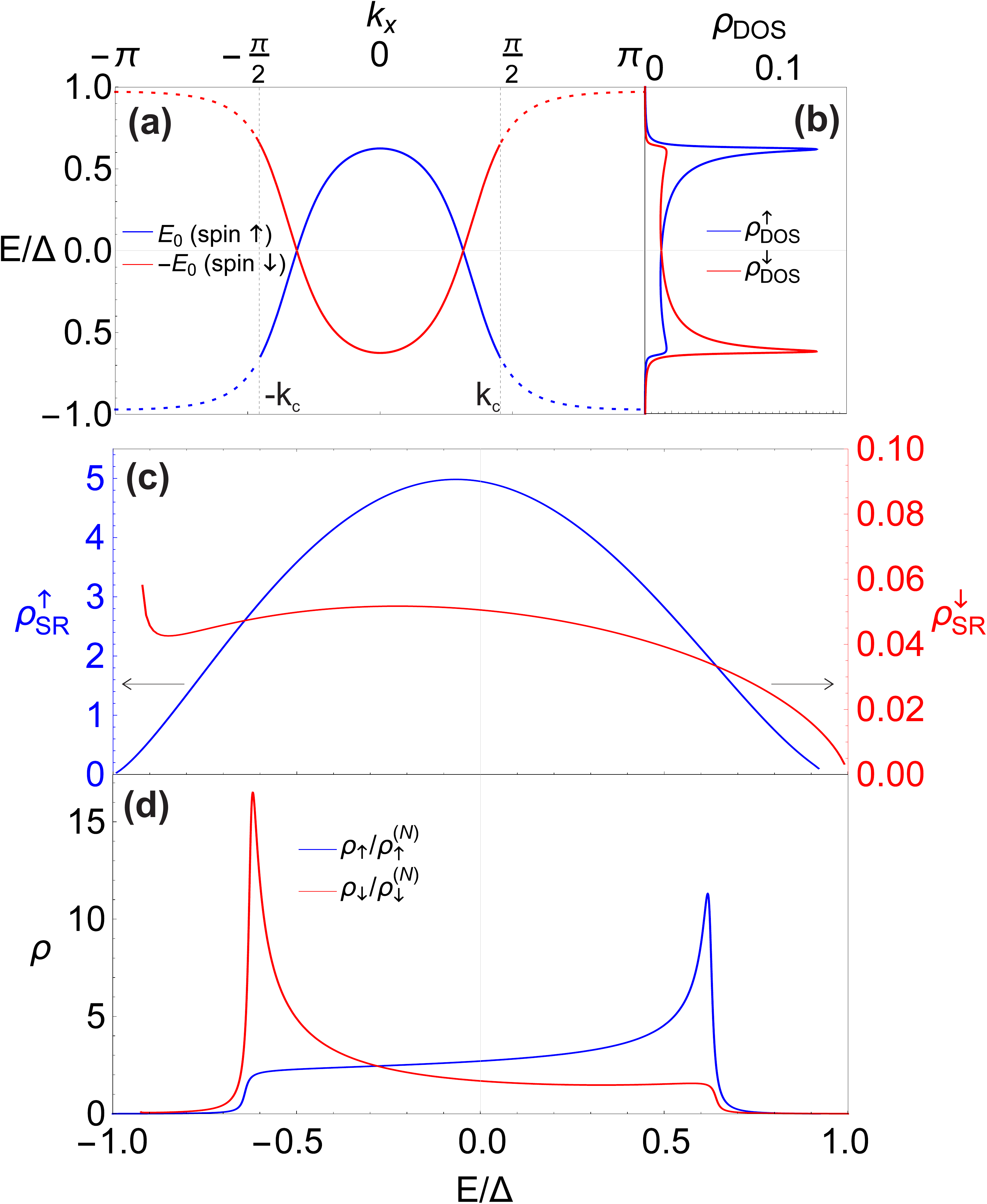}
  \caption{Shiba bands in the absence of spin-orbit coupling. (a) Dispersion relations from Eq.~\eqref{eq:E0_shiba_band} with $\xi_d(k_x) = 2t_d\cos k_x$, and (b) the corresponding density of states $\rho_{\text{DOS}}^{\uparrow/\downarrow}(E)$ with a momentum cutoff by the Fermi wave-vector $k_c$; (c) $\rho_{\text{SR}}^{\uparrow/\downarrow}$ [see Eq.~\eqref{eq:rho_sr} and text for definition] as a function of energy $E$; (d) $\rho_{\uparrow/\downarrow}$ as a product of $\rho_{\text{DOS}}^{\uparrow/\downarrow}$ and $\rho_{\text{SR}}^{\uparrow/\downarrow}$, scaled by $\rho_{\uparrow/\downarrow}^{(N)}$ [see Eqs.~\eqref{eq:rho_E} and \eqref{eq:rho_band_normal}], as a function of energy. Note that in the absence of spin-orbit coupling, the Shiba bands have definite spin polarizations (either $\uparrow$ or $\downarrow$) in terms of their electronic components. The parameters used here are: $M=1$, $\mu=1.2$, $v=0.2$, $t_d=0.2$, $\Delta=0.001$ and $k_c = 0.45\pi$.}
  \label{fig:rhos_band}
\end{figure}

Eq.~\eqref{eq:rho_band_org} can be further simplified by noticing that Eq.~\eqref{eq:E0_shiba_band} implies (assuming $\mu_d>0$)
\begin{align}
  \mu_d[k_i^{\pm}(E)] = \mu_d^{\pm}(E) = \sqrt{M^2 - v^2 \mp 2Mv_E}, \label{eq:muE}
\end{align}
with $v_E = vE/\sqrt{\Delta^2-E^2}$ as in Eq.~\eqref{eq:vE}. Namely, $\mu_d[k_i^{\pm}(E)]$ is only a function of $E$ (denoted by $\mu_d^{\pm}(E)$ henceforth) and is irrespective of $k_i$. Therefore $\rho_{\uparrow/\downarrow}(E)$ in Eq.~\eqref{eq:rho_band_org} can be factorized into two parts:
\begin{align}
  &\rho_{\uparrow/\downarrow}(E) = \rho_{\text{DOS}}^{\uparrow/\downarrow}(E)\cdot \rho_{\text{SR}}^{\uparrow/\downarrow}(E), \label{eq:rho_band}\\
  &\rho_{\text{DOS}}^{\uparrow/\downarrow}(E) =  \sum_{k_i^{\pm}} \left|\frac{1}{\partial E_0/\partial k_x}\right|_{k_x=k_i^{\pm}(E)}, \label{eq:rho_dos}\\
  &\rho_{\text{SR}}^{\uparrow/\downarrow}(E) = \frac{v\sqrt{\Delta^2 - E^2}}{[\mu_d^{\pm}(E) \mp M]^2 + v^2}. \label{eq:rho_sr}
\end{align}
Here, $\rho_{\text{DOS}}^{\uparrow/\downarrow}$ are the DOS of the Shiba bands; $\rho_{\text{SR}}^{\uparrow/\downarrow}$ are the spin densities inherited from the sum rule Eq.\eqref{eq:sum_rule0}. In essence, Eq.~\eqref{eq:rho_band} is the same as Eq.~\eqref{eq:rho_subgap_main} with the $\delta$-function replaced by the DOS. Note that the $E$-dependence in Eq.~\eqref{eq:rho_sr} corresponds to the $E_0$-dependence in Eq.\eqref{eq:sum_rule0}.

Owing to the particle-hole symmetry $k_i^+(+E) = k_i^-(-E)$, $\rho_{\text{DOS}}^{\uparrow/\downarrow}$ satisfy
\begin{align}
  \forall |E|<\Delta: \quad \rho_{\text{DOS}}^{\uparrow}(E) = \rho_{\text{DOS}}^{\downarrow}(-E). \label{eq:rho_dos_sym}
\end{align}
Moreover, $\rho_{\text{DOS}}^{\uparrow/\downarrow}$ contain Van Hove singularities of the Shiba bands that are dominant in the $E$-dependence of $\rho_{\uparrow/\downarrow}$. In Fig.~\ref{fig:rhos_band} (a) and (b), we show an example of the Shiba bands from Eq.~\eqref{eq:E0_shiba_band} with $\xi_d(k_x) = 2t_d \cos k_x$, and its corresponding $\rho_{\text{DOS}}^{\uparrow/\downarrow}$ with a specific $k_c$. On the other hand, $\rho_{\text{SR}}^{\uparrow/\downarrow}$ do not apparently exhibit any particle-hole symmetry [see Fig.~\ref{fig:rhos_band} (c)] as they both originate from the broadening of the pristine $d$-orbital bands. In particular, $\rho_{\text{SR}}^{\uparrow}$ and $\rho_{\text{SR}}^{\downarrow}$ can differ significantly in magnitude when the chemical potential is much closer to one of the spin bands (the minority spin, $\uparrow$), therefore the spectral density defined as a summation of $\rho_{\uparrow}$ and $\rho_{\downarrow}$ will be mostly dominated by the former and hence exhibit a strong asymmetry inside the gap. This is indeed the observation of spin-independent STM measurements \cite{nadj-perge_observation_2014, feldman_high-resolution_2017}. We will further see, in Sec.~\ref{sec:stm}, how spin-polarized STM measurements can effectively amplify the spin $\downarrow$ density by normalizing $\rho_{\uparrow/\downarrow}$ with their associated normal-state background $\rho_{\uparrow/\downarrow}^{(N)}$ [see Fig.~\ref{fig:rhos_band} (d)].

\begin{figure}
  \centering
  \includegraphics[width=0.45\textwidth]{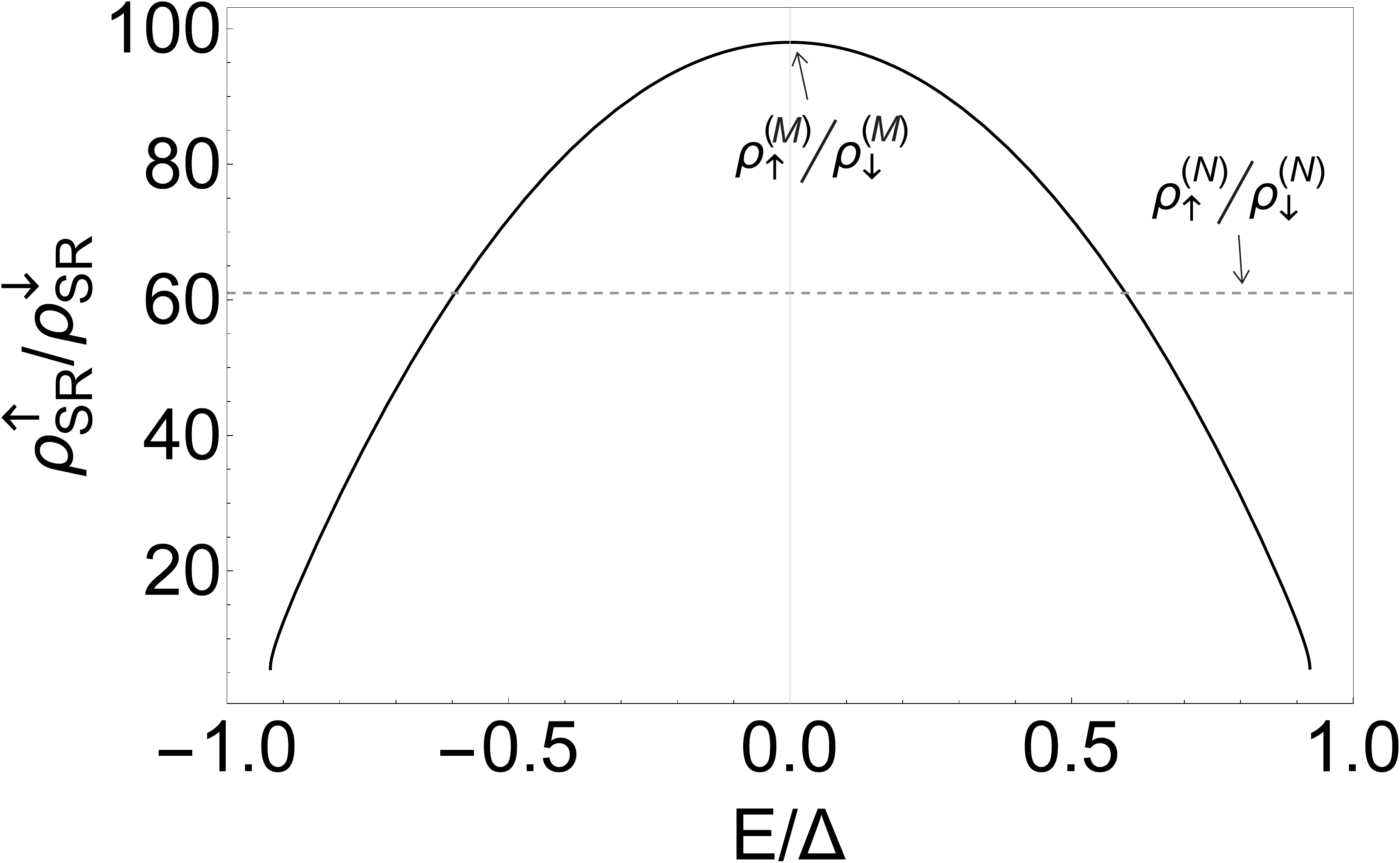}
  \caption{The ratio between $\rho_{\text{SR}}^{\uparrow}$ and $\rho_{\text{SR}}^{\downarrow}$ as a function of energy. The broken line marks the ratio of normal-state spin densities in the same parameter setting. The parameters used here are the same as in Fig.~\ref{fig:rhos_band}.}
  \label{fig:rhos_sr_ratio}
\end{figure}

Before we proceed, we point out one hidden symmetry in the ratio between $\rho_{\text{SR}}^{\uparrow}$ and $\rho_{\text{SR}}^{\downarrow}$. From Eqs.~\eqref{eq:rho_sr} and \eqref{eq:muE}, we have
\begin{align}
  &\frac{\rho_{\text{SR}}^{\uparrow}(E)}{\rho_{\text{SR}}^{\downarrow}(E)} =
  \frac{[\mu_d^{-}(E) + M]^2 + v^2}{[\mu_d^{+}(E) - M]^2 + v^2} \nn\\
  &=\frac{M+v_E+\mu_d^{-}(E)}{M-v_E-\mu_d^{+}(E)} \nn \\
  &=\frac{(M+v_E+\mu_d^{-}(E))(M-v_E+\mu_d^{+}(E))}{v^2\Delta^2/(\Delta^2-E^2)} \nn\\
  &=\frac{\rho_{\text{SR}}^{\uparrow}(-E)}{\rho_{\text{SR}}^{\downarrow}(-E)}. \label{eq:rho_SR_sym}
\end{align}
That is, the ratio $\rho_{\text{SR}}^{\uparrow}/\rho_{\text{SR}}^{\downarrow}$ is a symmetric function of $E$ (see Fig.~\ref{fig:rhos_sr_ratio}). This symmetry is in fact implied by Eq.~\eqref{eq:E0_shiba_main} as
\begin{align}
  1-(E_0/\Delta)^2 = 4 \pi^2 M^2 \rho_{\uparrow}^{(N)}\,\rho_{\downarrow}^{(N)},
\end{align}
which relates directly the Shiba state energy with the normal-state spin densities. In the context of Shiba bands, this relation becomes
\begin{align}
  \rho_{\text{SR}}^{\uparrow}(E)\,\rho_{\text{SR}}^{\downarrow}(-E) = \frac{(\Delta^2 - E^2)^2}{4M^2\Delta^2},
\end{align}
which immediately leads to the symmetry presented in Eq.~\eqref{eq:rho_SR_sym}.

More importantly, we find the maximum of the ratio $\rho_{\text{SR}}^{\uparrow}/\rho_{\text{SR}}^{\downarrow}$ at $E=0$, with
\begin{align}
  \hspace{-2mm}\max_E \left[\frac{\rho_{\text{SR}}^{\uparrow}(E)}{\rho_{\text{SR}}^{\downarrow}(E)}\right] = \frac{\rho_{\text{SR}}^{\uparrow}(E=0)}{\rho_{\text{SR}}^{\downarrow}(E=0)} =
  \frac{M+\sqrt{M^2 - v^2}}{M-\sqrt{M^2 - v^2}}. \label{eq:max_rhosr_ratio}
\end{align}
This is also the ratio of $\rho_{\uparrow}/\rho_{\downarrow}$ at zero energy as $\rho_{\text{DOS}}^{\uparrow}(E=0) = \rho_{\text{DOS}}^{\downarrow}(E=0)$ from Eq.~\eqref{eq:rho_dos_sym}. We will see that Majorana zero modes acquire precisely this maximum ratio. 

\subsection{Majorana zero modes with perturbative spin-orbit coupling}\label{ssec:Maj_pert}

We now investigate the spin densities associated with the Majorana zero modes by including spin-orbit coupling perturbatively. To this end we assume well-behaved $d$-orbital bands such that there exist and only exist two solutions, $\pm k_0$, to the equation $E_0(k_x) = 0$ in the limit of vanishing spin-orbit coupling. Namely,
\begin{align}
  M^2-\mu_d(\pm k_0)^2-v^2 = 0. \label{eq:k0def}
\end{align}
Here we have used the fact that $\xi_d(k_x)$ in Eq.~\eqref{eq:Hd} is an even function of $k_x$. By solving the effective Hamiltonian in the vicinity of $\pm k_0$ (see Appendix \ref{app:maj}), we obtain the $d$-orbital components of the Majorana zero modes to be (up to a normalization factor; note that the Majorana wavefunctions have support both in the magnetic chain and in the superconductor, but here we focus on the chain part only)
\begin{subequations}
\label{eq:wf_maj_chain}
\begin{align}
  &{\chi}_1(x) =
  \begin{pmatrix}
    \cos\frac{\theta_0}{2} \\ \sin\frac{\theta_0}{2} \\ \sin\frac{\theta_0}{2} \\ -\cos\frac{\theta_0}{2}
  \end{pmatrix} e^{- x/\lambda}\sin k_0 x,\quad (x/\lambda>0);\\
  &{\chi}_2(x) = i
  \begin{pmatrix}
    \cos\frac{\theta_0}{2} \\ -\sin\frac{\theta_0}{2} \\ \sin\frac{\theta_0}{2} \\ \cos\frac{\theta_0}{2}
  \end{pmatrix} e^{x/\lambda}\sin k_0 x,\quad (x/\lambda<0),
\end{align}
\end{subequations}
where
\begin{align}
  &\sin\theta_0 = v/M,\quad \cos\theta_0 = \sqrt{M^2-v^2}/M, \\
  &\lambda = \frac{\sqrt{M^2 - v^2}}{\xi_{SO}(k_0) v}\,\frac{\partial \xi_d}{\partial k_x}\Bigr|_{k_x=k_0}. \label{eq:lambda}
\end{align}
$\chi_1$ and $\chi_2$ correspond to two Majorana zero modes at two ends of the chain. The spinor parts of these wavefunctions, which are position-independent, are precisely given by the artificial Majorana solutions Eq.~\eqref{eq:majorana_toy} in the toy model with $v$ replacing $\Delta_d$ and $\varphi=0$.
The ratio of Majorana spin densities $\rho_{\uparrow/\downarrow}^{(M)}$, for both $\chi_1$ and $\chi_2$, is given by
\begin{align}
  &\frac{\rho_{\uparrow}^{(M)}}{\rho_{\downarrow}^{(M)}} =
  \frac{M+\sqrt{M^2-v^2}}{M-\sqrt{M^2-v^2}}, \label{eq:rhoM_ratio}
\end{align}
which echoes the maximum ratio in Eq.~\eqref{eq:max_rhosr_ratio}. In addition, from the wavefunctions Eq.~\eqref{eq:wf_maj_chain}, we have
\begin{subequations}
\label{eq:rhoM_decay}
\begin{align}
  &\rho_{\uparrow}^{(M)}(x) \propto \cos^2\frac{\theta_0}{2}e^{- 2|x/\lambda|}\sin^2 k_0 x,\\ 
  &\rho_{\downarrow}^{(M)}(x) \propto \sin^2\frac{\theta_0}{2}e^{- 2|x/\lambda|}\sin^2 k_0 x.
\end{align}
\end{subequations}

\subsection{Effects of finite spin-orbit coupling}

The case of finite spin-orbit coupling can be solved directly from Eqs.~\eqref{eq:Gd0}, \eqref{eq:rho_def} and \eqref{eq:Hd}, although the analytical expressions in general become lengthier and less transparent compared with the vanishing spin-orbit coupling case. We will focus on the spin densities associated with the Majorana zero modes in this case, and discuss briefly the spin densities associated with the Shiba bands at the end of this section.

\begin{figure}
  \centering
  \includegraphics[width=0.45\textwidth]{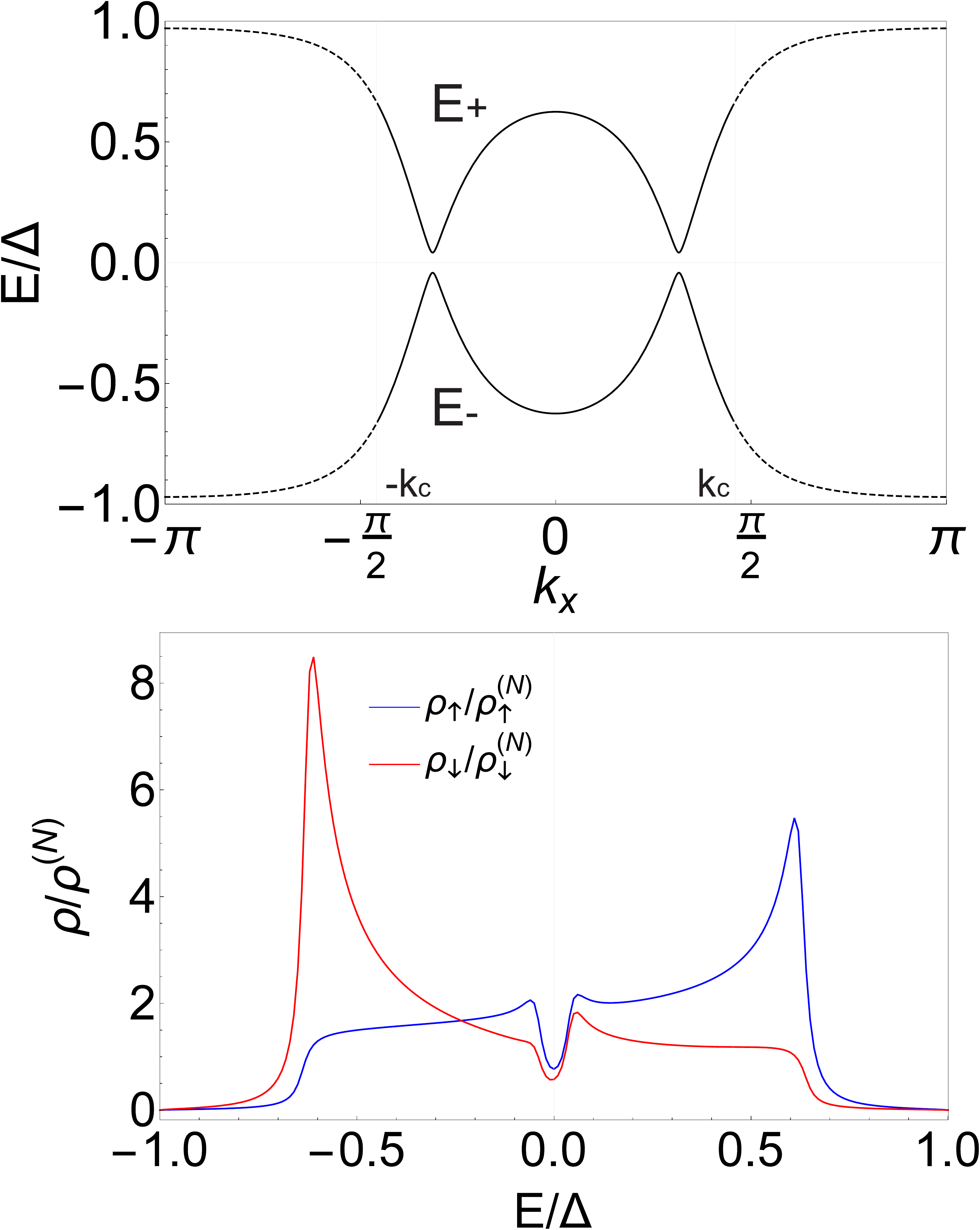}
  \caption{Shiba band dispersion relations (the upper panel), from Eq.~\eqref{eq:Epm_shiba_band}, and spin densities (the lower panel) normalized by the normal-state background given by Eqs.~\eqref{eq:rho_E} and \eqref{eq:rho_N_so}. Here we have assumed $\xi_{SO}(k_x) = \alpha\sin k_x$ and $\xi_d(k_x) = 2t_d\cos k_x$. The parameters used in this example are the same as those in Fig.~\ref{fig:rhos_band}, except for a finite spin-orbit coupling $\alpha = 0.05$.}
  \label{fig:shiba_band_so}
\end{figure}

In the presence of finite spin-orbit coupling, the Shiba band dispersion relations are given by (see Fig.~\ref{fig:shiba_band_so} upper panel, and Appendix \ref{app:soc})
\begin{align}
  E_{\pm} \simeq \pm\Delta\sqrt{\frac{[(M^2+\xi_{SO}^2)-(\mu_d^2+v^2)]^2+4\xi_{SO}^2 v^2}{[(M^2+\xi_{SO}^2)-(\mu_d^2+v^2)]^2+4(M^2+\xi_{SO}^2)v^2}}, \label{eq:Epm_shiba_band}
\end{align}
where we have dropped the $k_x$ dependence of $E_{\pm}$, $\mu_d$ and $\xi_{SO}$ to shorten the expression. This equation is to be compared with Eq.~\eqref{eq:E0_shiba_band}. Clearly, $E_{\pm}(k_x)|_{\xi_{SO}\rightarrow0} = \pm |E_0(k_x)|$. The solutions of $E_{\pm}=0$ exist only if $\xi_{SO}=0$ and $M^2=\mu_d^2+v^2$, the former generically requiring $k_x = 0$ or $\pi$, and the latter imposing in addition a condition for the values of $\xi_d(k_x)$ (and hence $\mu_d(k_x)$) at these special momenta. When fully gapped (see Fig.~\ref{fig:shiba_band_so} for an example), the Shiba bands are topologically nontrivial if (see Appendix \ref{app:soc})
\begin{align}
  \mathrm{sgn}[M^2 - \mu_d(0)^2 - v^2]\cdot\mathrm{sgn}[M^2 - \mu_d(\pi)^2 - v^2] = -1. \label{eq:topo_cond_main}
\end{align}
If this condition is true, the induced $p$-wave gap $\Delta_p$ estimated at $k_0$, where by definition
\begin{align}
  M^2+\xi_{SO}(\pm k_0)^2-\mu_d(\pm k_0)^2-v^2 = 0, \label{eq:k0def_so}
\end{align}
is given by
\begin{align}
  \Delta_{p} = |E_{\pm}(k_0)| = \Delta\sqrt{\frac{\xi_{SO}(k_0)^2}{M^2+\xi_{SO}(k_0)^2}}.
\end{align}
If $M\gg |\xi_{SO}(k_0)|$, then $\Delta_{p}\approx{\Delta}|\xi_{SO}(k_0)|/M$, which recovers the estimation given by Ref.~\onlinecite{li_topological_2014} (see also Eq.~\eqref{eq:Delta_p_pert} in Appendix \ref{app:maj}). Furthermore, the Majorana zero mode solutions are given by (up to a normalization factor; see Appendix \ref{app:soc})
\begin{subequations}\label{eq:wf_maj_so}
\begin{align}
  \chi_1(x)
  &=
  \begin{pmatrix}
    \chi_{\uparrow}(x) \\
    \chi_{\downarrow}(x) \\
    \chi_{\downarrow}(x) \\
    -\chi_{\uparrow}(x)
  \end{pmatrix} e^{-x/\lambda}, \quad (x/\lambda>0) \\
  \chi_2(x)
  &= i
  \begin{pmatrix}
    \chi_{\uparrow}(-x) \\
    -\chi_{\downarrow}(-x) \\
    \chi_{\downarrow}(-x) \\
    \chi_{\uparrow}(-x)
  \end{pmatrix} e^{x/\lambda}, \quad (x/\lambda<0)
\end{align}
\end{subequations}
where
\begin{align}
  &\chi_{\uparrow}(x) \simeq [M+{\mu}_d(k_0)]\sin k_0 x - \frac{\xi_{SO}(k_0)v}{{\mu}_d(k_0)}\cos k_0 x, \\
  &\chi_{\downarrow}(x) \simeq v\sin k_0 x + \xi_{SO}(k_0)\cos k_0 x, \\
  &\lambda \simeq \frac{{\mu}_d(k_0)}{\xi_{SO}(k_0) v}\,\frac{\partial \xi_d}{\partial k_x}\Bigr|_{k_x=k_0}, \label{eq:lambda_so}
\end{align}
and we have assumed $k_0 \lambda \gg 1$ in the above approximate expressions. Note that we have intentionally reused several notations, $k_0$, $\lambda$ and $\chi_{1,2}$, that have appeared in Sec.~\ref{ssec:Maj_pert}, because the same notations share exactly the same physical meaning, and they become equivalent in the small spin-orbit coupling limit [cf. Eqs.~\eqref{eq:k0def_so}, \eqref{eq:lambda_so} and \eqref{eq:wf_maj_so} versus Eqs.~\eqref{eq:k0def}, \eqref{eq:lambda} and \eqref{eq:wf_maj_chain}].

From Eq.~\eqref{eq:wf_maj_so}, the ratio of the Majorana spin densities, ${\rho_{\uparrow}^{(M)}(x)}/{\rho_{\downarrow}^{(M)}(x)} \equiv |{\chi_{\uparrow}(x)}/{\chi_{\downarrow}(x)}|^2$, is no longer $x$-independent [cf. Eq.~\eqref{eq:rhoM_ratio}] because of the spin precession induced by the finite spin-orbit coupling. For practical purposes we define a ratio of the integrated Majorana spin densities as
\begin{align}
  \frac{\rho_{\uparrow}^{(M)}}{\rho_{\downarrow}^{(M)}}
  &\equiv
  \frac{\int_{0}^{{\pi}/{k_0}}dx\,|{\chi_{\uparrow}(x)}|^2}{\int_{0}^{{\pi}/{k_0}}dx\,|{\chi_{\downarrow}(x)}|^2} \\
  &=\frac{[M+{\mu}_d(k_0)]^2+[{\xi_{SO}(k_0)v}/{{\mu}_d(k_0)}]^2}{v^2+\xi_{SO}(k_0)^2}. \label{eq:rhoM_ratio_so}
\end{align}
When $\xi_{SO}(k_0)$ is negligible, this ratio becomes Eq.~\eqref{eq:rhoM_ratio} by using Eq.~\eqref{eq:k0def_so}; otherwise this ratio is reduced in magnitude under realistic conditions $[M+{\mu}_d(k_0)]/v \gg v/{{\mu}_d(k_0)}$. As we will see in the next section, it is important to compare this ratio with that of the normal-state spin densities, which can be obtained straightforwardly by setting $\Delta=0$ in Eq.~\eqref{eq:Gd0} and by including the $k_x$ dependence such that $G_d(k_x, E^+) \simeq [E - H_d(k_x) + iv]^{-1}$. We have
\begin{align}
  &\rho_{\uparrow/\downarrow}^{(N)}(k_x, E) \nn\\
  &\;= \frac{[(E+\mu_d \pm M)^2+\xi_{SO}^2+v^2]v/\pi}{[M^2+\xi_{SO}^2-(E+\mu_d)^2-v^2]^2+4(M^2+\xi_{SO}^2)v^2}, \label{eq:rho_N_so}
\end{align}
and
\begin{align}
  \frac{\rho_{\uparrow}^{(N)}(k_x, E=0)}{\rho_{\downarrow}^{(N)}(k_x, E=0)}
  = \frac{(\mu_d + M)^2+\xi_{SO}^2+v^2}{(\mu_d - M)^2+\xi_{SO}^2+v^2}, \label{eq:rhoN_ratio_so}
\end{align}
where both $\mu_d$ and $\xi_{SO}$ are functions of $k_x$. We will postpone a detailed discussion about the comparison between Majorana and normal-state spin densities to the next section where its physical implication in STM measurements becomes clear.

\begin{figure}
  \centering
  \includegraphics[width=0.45\textwidth]{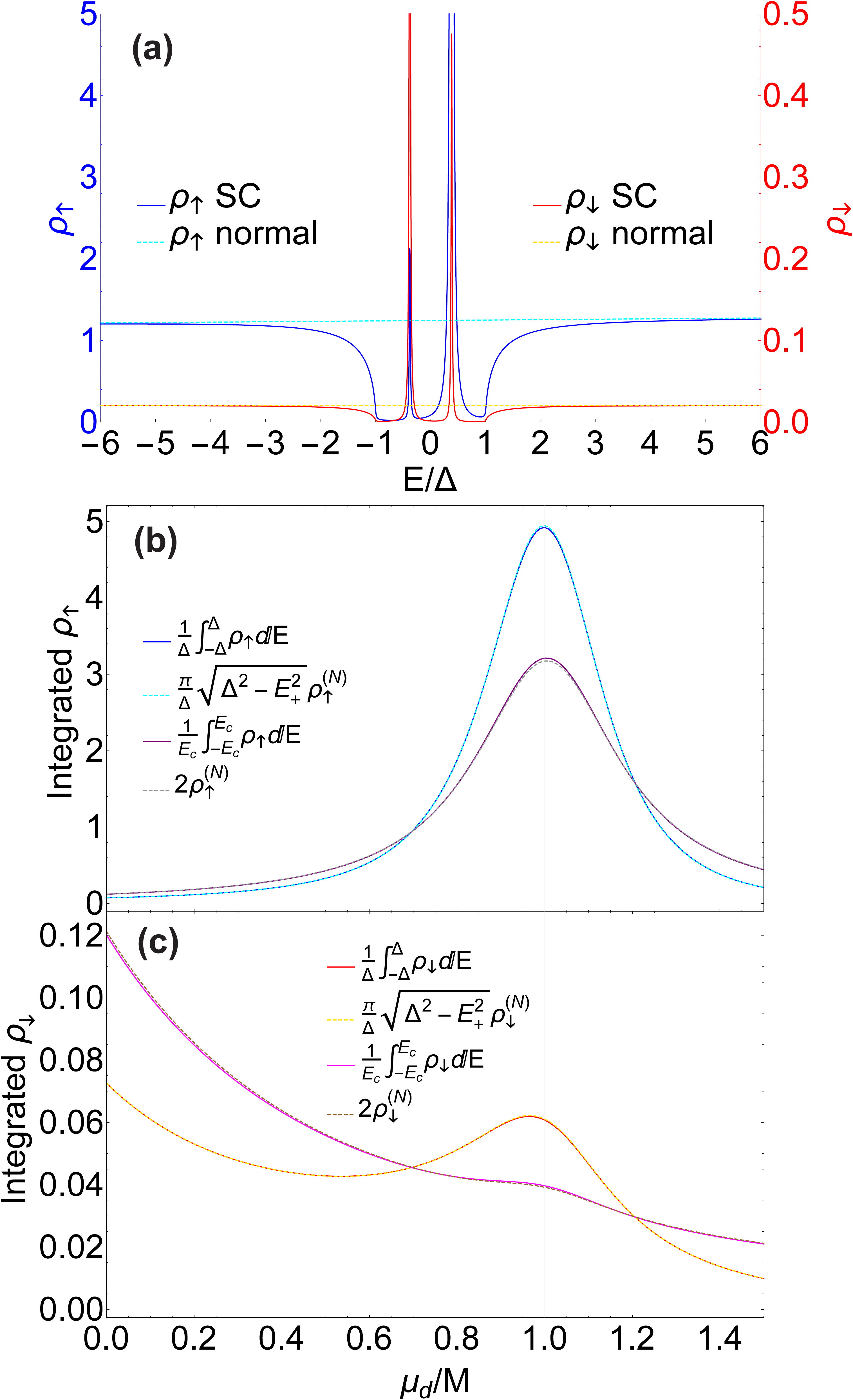}
  \caption{Spin densities in a magnetic impurity chain at a specific $k_x$ with generic $\mu_d$ and $\xi_{SO}$. (a) The energy dependence of the spin densities in the superconducting state (solid lines), obtained directly from Eqs.~\eqref{eq:Gd0}, \eqref{eq:rho_def} and Eq.~\eqref{eq:Hd}, and in the normal state (broken lines), obtained from Eq.~\eqref{eq:rho_N_so}; (b) and (c), integrated spin densities ($\uparrow$ in (b) and $\downarrow$ in (c)) in $[-\Delta, \Delta]$ and $[-E_c, E_c]$ with $E_c = 6\Delta$, as functions of $\mu_d$. Panels (b) and (c) verifies both the sum rule Eq.~\eqref{eq:sum_rule_so} and \eqref{eq:sum_rule_so_full}. The parameters used in these plots are $M=1$, $v=0.2$, $\xi_{SO}=0.1$, $\Delta=0.001$ and $\eta=1\mathrm{e}-5$ (cf. Fig.~\ref{fig:rhos_single}). Additionally, for panel (a), $\mu_d=0.9$.}
  \label{fig:rhos_singlek_so}
\end{figure}

Now we turn our attention to the spin densities associated with the Shiba bands whose dispersion relations are given by Eq.~\eqref{eq:Epm_shiba_band}. By directly solving Eqs.~\eqref{eq:Gd0} and \eqref{eq:rho_def} with the Hamiltonian in Eq.~\eqref{eq:Hd}, we obtain (see Appendix \ref{app:soc})
\begin{align}
  &\rho_{\uparrow/\downarrow}(k_x, E) \nn\\
  &\quad\simeq \frac{\pi}{2} \sqrt{\Delta^2 - E^2}\, \left[\rho_{\uparrow/\downarrow}^{(N)}(k_x, E=0) \pm \mathrm{sgn}(E)\rho_{\uparrow/\downarrow}^{(A)}(k_x)\right] \nn\\
  &\qquad\cdot\Bigl\{\delta\bigl[E - E_+(k_x)\bigr]+\delta\bigl[E - E_-(k_x)\bigr]\Bigr\}, \quad (|E|<\Delta)\label{eq:rho_subgap_so}
\end{align}
where $\rho_{\uparrow/\downarrow}^{(N)}(k_x, E=0)$ are the normal-state spin densities, given by Eq.~\eqref{eq:rho_N_so}, at $E=0$, and (suppressing the $k_x$-dependence)
\begin{align}
  &\rho_{\uparrow/\downarrow}^{(A)} = \frac{|Z|v/\pi}{|Z|^2 + 4M^2v^2}\,
  \left[2M(M\pm\mu_d)\mathrm{Re}\left(\frac{1}{Z}\right) - 1\right], \label{eq:rho_A} \\
  &Z \equiv M^2+\xi_{SO}^2 - \mu_d^2 - v^2 + 2i\xi_{SO} v.
\end{align}
At $k_x$ where $\xi_{SO}(k_x)=0$, it is straightforward to verify that $\rho_{\uparrow/\downarrow}^{(A)}(k_x) = \mathrm{sgn}[E_0(k_x)]\rho_{\uparrow/\downarrow}^{(N)}(k_x,E=0)$ and $E_{\pm}(k_x) = \pm |E_0(k_x)|$, with $E_0(k_x)$ given by Eq.~\eqref{eq:E0_shiba_band}. Hence Eq.~\eqref{eq:rho_subgap_so} becomes Eq.~\eqref{eq:rho_subgap_main} with $\mu$ replaced by $\mu_d(k_x)$. If $\xi_{SO}(k_x)\ne 0$, on the other hand, the spin densities are generically distributed into two delta functions at opposite energies $E=E_{\pm}(k_x)$ with asymmetric weights represented by $\rho_{\uparrow/\downarrow}^{(A)}$ [see Fig.~\ref{fig:rhos_singlek_so}(a)].

For each $k_x$, a sum rule similar to Eq.~\eqref{eq:sum_rule0} holds true even in the presence of finite spin-orbit coupling [see Fig.~\ref{fig:rhos_singlek_so}(b) and (c)], as from Eq.~\eqref{eq:rho_subgap_so} we have
\begin{align}
  \int_{-\Delta}^{\Delta}\rho_{\uparrow/\downarrow}(k_x, E)\, dE \simeq \pi\sqrt{\Delta^2 - E_+(k_x)^2}\; \rho_{\uparrow/\downarrow}^{(N)}(k_x, E=0). \label{eq:sum_rule_so}
\end{align}
We have also verified numerically that another sum rule similar to Eq.~\eqref{eq:sum_rule_full} holds true for each $k_x$ (see Fig.~\ref{fig:rhos_singlek_so}(b) and (c)):
\begin{align}
  \int_{-E_c}^{E_c}\rho_{\uparrow/\downarrow}(k_x, E)\, dE \simeq 2E_c\,\rho_{\uparrow/\downarrow}^{(N)}(k_x, E=0), \label{eq:sum_rule_so_full}
\end{align}
where $\Delta\ll E_c\ll v,M$. In terms of the integrated spin densities $\rho_{\uparrow/\downarrow}(E)$ defined in Eq.~\eqref{eq:rho_E}, however, the factorization into a DOS part and a sum-rule part as in Eq.~\eqref{eq:rho_band} becomes spoiled in general because of the separation of the subgap spin densities to different energies. This makes an analytical treatment of $\rho_{\uparrow/\downarrow}(E)$ intractable in the presence of generic spin-orbit coupling. Nevertheless, we emphasize that, since Van Hove singularities of Shiba bands are likely to occur at $k_x = 0$ or $\pi$ owing to symmetry (see Fig.~\ref{fig:shiba_band_so} lower panel, for example), the major experimental features associated with Shiba-band spin densities will be dominated by the states around these special momenta where spin-orbit coupling is negligible, hence we expect our analysis in Sec.~\ref{ssec:shiba_band} to remain a good account for major experimental features. In Sec.~\ref{sec:sim}, we will show numerical simulations that fully take account of non-perturbative spin-orbit coupling.

\section{STM measurements}\label{sec:stm}

In this section, we formulate a phenomenological theory that captures the key ingredients of spin-polarized STM measurements. Using this theory, as well as the results obtained in the previous sections, to understand the key features observed in recent experiments \cite{Jeong_2017_to_appear}.

\subsection{General theory}

We consider two possible directions of the tip spin polarization, denoted by N and P. We assume, according to Fermi's golden rule, the tunneling current measured at a specific tip position $\bm{r}$ and a specific bias voltage $V$ is associated with the local spin densities $\rho_{\uparrow/\downarrow}(\bm{r}', E)$ as follows:
\begin{align}
  &I_{N/P}(\bm{r}, V) \nn\\
  &\;= \int_0^{eV} dE \int d\bm{r}'\; \sum_{\sigma=\uparrow,\downarrow}w_{N/P,\sigma}(\bm{r}-\bm{r}')\,\rho_{\sigma}(\bm{r}', E), \label{eq:I_general}
\end{align}
where $w_{N/P,\uparrow/\downarrow}$ are non-negative weight factors that are assumed to be energy independent. In the following analysis, we consider the spin densities associated with the $d$-orbital electrons of the magnetic chain (at $y'=z'=0$ in Eq.~\eqref{eq:I_general}) and consider only measurements along the chain ($y=0$ in Eq.~\eqref{eq:I_general}); we will also assume the weight factors to be proportional to $\delta(x-x')$ along the chain. More generic and realistic conditions will be considered in numerical simulations that will be presented in the next section. With the preceding constraint, Eq.~\eqref{eq:I_general} is simplified to
\begin{align}
  &I_{N/P}(x,z,V) = \sum_{\sigma=\uparrow,\downarrow}w_{N/P,\sigma}(z) \int_0^{eV} dE\; \rho_{\sigma}(x, E). \label{eq:I_chain}
\end{align}
Physically, the weight factors contain a contribution associated with the relative angle between the spin polarization of the tip (N and P) and that of the chain ($\uparrow$ and $\downarrow$), as well as a contribution associated with the spatial dependence of the electronic wavefunctions in the tip and in the chain. We will assume that these two contributions are separable as $w_{N/P,\sigma}(z) = w'_{N/P,\sigma} w''(z)$. As a consequence, the ratio of any two of these weight factors is independent on $z$.

In actual STM measurements, the height of the tip ($z$) is set for each specific $x$ by keeping the total current measured at a particular bias $V_c = E_c/e$ to be a constant. This so-called set-point effect, combined with the assumption of separability of the weight factors, leads to a normalization factor to the measured differential conductance such that (see Appendix \ref{app:cond})
\begin{align}
  G_{N/P}(x,E) &= \frac{\tilde{w}_{N/P}\,\rho_{\uparrow}(x,E)+\rho_{\downarrow}(x,E)}{\tilde{w}_{N/P}\,R_{\uparrow}(x)+R_{\downarrow}(x)}, \label{eq:Gmodel}
\end{align}
where
\begin{align}
  &R_{\uparrow/\downarrow}(x) = \int_0^{E_c} dE\; \rho_{\uparrow/\downarrow}(x,E), \label{eq:Rud} \\
  &\tilde{w}_{N/P} = {w_{N/P,\uparrow}}/{w_{N/P,\downarrow}}. \label{eq:tildew}
\end{align}
Note that $\tilde{w}_{N/P}$ do not depend on $x$, $z$ or $E$. It follows that the difference between the conductances measured with two tip polarizations is given by
\begin{align}
  \delta G = G_N-G_P = \frac{(\tilde{w}_N - \tilde{w}_P)(\tilde{\rho}_{\uparrow} - \tilde{\rho}_\downarrow)}{(\tilde{w}_N R_{\uparrow}/R_{\downarrow} + 1)(\tilde{w}_P + R_{\downarrow}/R_{\uparrow})}, \label{eq:dG}
\end{align}
where we have dropped the $x$ and/or $E$ dependence of the variables to shorten the expression, and we have defined normalized spin densities
\begin{align}
  &\tilde{\rho}_{\uparrow/\downarrow}(x, E) = \frac{\rho_{\uparrow/\downarrow}(x,E)}{R_{\uparrow/\downarrow}(x)}.
\end{align}
Clearly, $\delta G$ vanishes whenever $\tilde{\rho}_{\uparrow} = \tilde{\rho}_\downarrow$. Moreover, if $\tilde{\rho}_{\uparrow}(x, E) = \tilde{\rho}_{\downarrow}(x, -E)$, then $\delta G(x, E) = -\delta G(x, -E)$. In what follows, we will assume $\tilde{w}_N - \tilde{w}_P > 0$ without loss of generality.

\subsection{Measurements on magnetic impurities}

We now proceed to show the implication of Eq.~\eqref{eq:dG} when we apply the results obtained in the previous sections, with the experimentally relevant condition $\Delta \ll E_c \ll v,M$. This condition allows us to approximate, by Eq.~\eqref{eq:sum_rule_so_full},
\begin{align}
  R_{\uparrow/\downarrow}(x) \simeq E_c\, \rho_{\uparrow/\downarrow}^{(N)}(x, E=0), \label{eq:R_to_rho_normal}
\end{align}
where $\rho_{\uparrow/\downarrow}^{(N)}(x, E=0)$ are the normal-state spin densities. We will focus on the subgap regime in what follows.

Let us first consider the case of a single magnetic impurity ($x$ will be dropped). In this case we use Eq.~\eqref{eq:rho_subgap_main} and take into account a finite broadening of the spectrum by replacing the delta function $\delta(E)$ with a symmetric function $f(E)$ satisfying $f(E) = f(-E)$. Then the normalized spin densities become
\begin{align}
  \tilde{\rho}_{\uparrow/\downarrow}(E) = \frac{\pi\sqrt{\Delta^2 - E_0^2}}{E_c} \;f(E \mp E_0). \label{eq:normalized_rho}
\end{align}
We immediately find $\tilde{\rho}_{\uparrow}(E) = \tilde{\rho}_{\downarrow}(-E)$, and hence
\begin{align}
  \delta G(E) = -\delta G(-E) \quad(\textit{single magnetic impurity}). \label{eq:deltaG_sym}
\end{align}
In particular, $\delta G$ vanishes at zero energy.

Next we consider the case of a magnetic impurity chain with a Majorana zero mode at its end. In this case we are most interested in $\delta G(x\rightarrow 0, E=0)$ and therefore a comparison between $\tilde{\rho}_{\uparrow}$ and $\tilde{\rho}_{\downarrow}$ at $x\rightarrow 0, E=0$. This comparison can be translated to comparing $({\rho}_{\uparrow}/{\rho}_{\downarrow})_{x\rightarrow 0, E=0}$ and $R_{\uparrow}/R_{\downarrow}$, where
\begin{align}
  R_{\uparrow/\downarrow} = E_c\,\int_{-k_c}^{k_c} dk_x\, \rho_{\uparrow/\downarrow}^{(N)}(k_x, E=0), \label{eq:R_band_normal}
\end{align}
by using Eq.~\eqref{eq:R_to_rho_normal} and by assuming $R_{\uparrow/\downarrow}$ to be independent on $x$.

We start with the vanishing spin-orbit coupling limit, where $\rho_{\uparrow/\downarrow}^{(N)}(k_x, E=0)$ is given by the integrand in Eq.~\eqref{eq:rho_band_normal} with $E=0$. At the end of the chain in the presence of a Majorana zero mode, we have, from Eq.~\eqref{eq:rhoM_ratio},
\begin{align}
  &\left(\frac{\rho_{\uparrow}}{\rho_{\downarrow}}\right)_{x\rightarrow 0, E=0} = \frac{\rho_{\uparrow}^{(M)}}{\rho_{\downarrow}^{(M)}}  =
  \frac{M+\sqrt{M^2-v^2}}{M-\sqrt{M^2-v^2}} \approx \frac{4M^2}{v^2},
\end{align}
where we have used the realistic assumption $v^2/M^2\ll 1 $ and kept only the leading order term with respect to $v^2/M^2$ in the final expression. On the other hand, from Eq.~\eqref{eq:R_band_normal}, we have
\begin{align}
  &\frac{R_{\uparrow}}{R_{\downarrow}} = \frac{\rho^{(N)}_{\uparrow}}{\rho^{(N)}_{\downarrow}} = \frac{\int_{-k_c}^{k_c} dk_x\, \rho_{\uparrow}^{(N)}(k_x, E=0)}{\int_{-k_c}^{k_c} dk_x\, \rho_{\downarrow}^{(N)}(k_x, E=0)} \label{eq:ratio_normal} \\
  &= \frac{\int_{-k_c}^{k_c} dk_x\, \frac{v / \pi}{[\mu_d(k_x) - M]^2 + v^2}}{\int_{-k_c}^{k_c} dk_x\, \frac{v / \pi}{[\mu_d(k_x) + M]^2 + v^2}} \nn \\
  &\quad< \mathrm{sup}\left\{\frac{(\mu + M)^2 + v^2}{(\mu - M)^2 + v^2} \: : \: \mu\in\Bigl\{\mu_d(k_x)\: : \: |k_x|\le k_c\Bigr\}\right\} \nn \\
  &\quad\le \frac{\sqrt{M^2 + v^2} + M}{\sqrt{M^2 + v^2} - M}
  \approx \frac{4M^2}{v^2}, \label{eq:R_ratio_ineq}
\end{align}
where we have again kept only the leading term in $v^2/M^2$ in the last step. Therefore ${\rho_{\uparrow}^{(M)}}/{\rho_{\downarrow}^{(M)}} > {R_{\uparrow}}/{R_{\downarrow}}$ in the limit of vanishing spin-orbit coupling. In Fig.~\ref{fig:rhos_sr_ratio}, we show one typical example of the comparison between ${\rho_{\uparrow}^{(M)}}/{\rho_{\downarrow}^{(M)}}$, which corresponds to the maximum point of the solid curve [see Eqs.~\eqref{eq:max_rhosr_ratio} and \eqref{eq:rhoM_ratio}], and ${R_{\uparrow}}/{R_{\downarrow}}$, which corresponds to the dashed flat line. Physically, the Majorana ratio ${\rho_{\uparrow}^{(M)}}/{\rho_{\downarrow}^{(M)}} \simeq {4M^2}/{v^2}$ is the ratio of the broadened ($\sim v$) normal-state spin densities at the Fermi energy, when the chemical potential is aligned with the pristine spin-$\uparrow$ level and the two pristine spin levels are separated by the exchange energy $2M$ (cf. $\rho_{\uparrow/\downarrow}^{(N)}(E)$ in Eq.~\eqref{eq:rho_normal_main} with $E=0$ and $\mu=M$). The reason for this is that Majorana zero modes originate from the opening of the induced $p$-wave gap which only concerns pristine $d$-orbital states that are very close to the Fermi energy (together with their pairing partners). In other words, Majorana zero modes are intimately associated with those states with $k_x$ such that $\mu_d(k_x) \approx M$ -- these are also the states that exhibit the maximum spin polarization $\rho_{\uparrow}^{(N)}/\rho_{\downarrow}^{(N)}$ at the Fermi energy. In contrast, the background ratio ${R_{\uparrow}}/{R_{\downarrow}}$ involves an average over a large range of $k_x$ determined by $k_c$, and hence is necessarily significantly smaller than ${\rho_{\uparrow}^{(M)}}/{\rho_{\downarrow}^{(M)}}$ in a generic setting.

\begin{figure}
  \centering
  \includegraphics[width=0.45\textwidth]{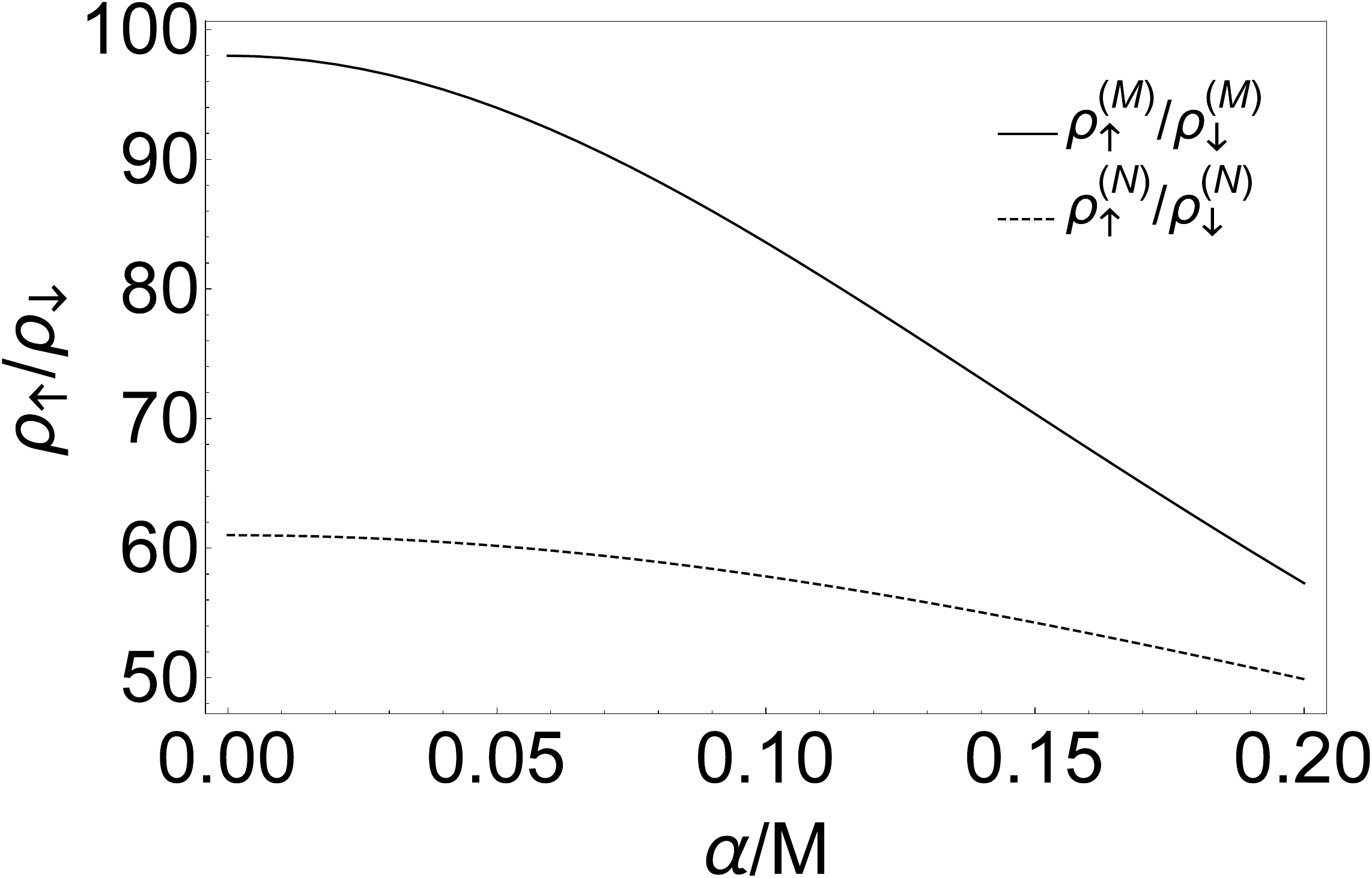}
  \caption{Comparison between the Majorana ratio $\rho^{(M)}_{\uparrow}/\rho^{(M)}_{\downarrow}$, given by Eq.~\eqref{eq:rhoM_ratio_so}, and the normal-state ratio $\rho^{(N)}_{\uparrow}/\rho^{(N)}_{\downarrow}$, given by Eqs.~\eqref{eq:rho_N_so} and \eqref{eq:ratio_normal}, both as a function of spin-orbit coupling parameter $\alpha$. Here we have assumed $\xi_{SO}(k_x) = \alpha\sin k_x$ and $\xi_d(k_x) = 2t_d\cos k_x$. Note that $\rho^{(N)}_{\uparrow}/\rho^{(N)}_{\downarrow}={R_{\uparrow}}/{R_{\downarrow}}$ owing to Eq.~\eqref{eq:R_band_normal}. The parameters used here are the same as in Fig.~\ref{fig:rhos_band} or \ref{fig:shiba_band_so} except for $\alpha$.}
  \label{fig:ratio_compare}
\end{figure}

\subsection{Effects of finite spin-orbit coupling on measurements}

When finite spin-orbit coupling is taken into account, the Majorana spin polarization is given by Eq.~\eqref{eq:rhoM_ratio_so} and the corresponding ($k_x$-dependent) normal-state spin polarization is given by Eq.~\eqref{eq:rhoN_ratio_so}. With the realistic assumption $\xi_{SO}^2, v^2 \ll M^2, \mu_d^2$, Eqs.~\eqref{eq:rhoM_ratio_so} and \eqref{eq:rhoN_ratio_so} can be approximated by
\begin{align}
  &\frac{\rho_{\uparrow}^{(M)}}{\rho_{\downarrow}^{(M)}}
  \approx \frac{4M^2}{\xi_{SO}(k_0)^2+v^2}, \label{eq:rhoM_ratio_so_approx} \\
  &\frac{\rho_{\uparrow}^{(N)}(k_x, E=0)}{\rho_{\downarrow}^{(N)}(k_x, E=0)}
  \approx \frac{[\mu_d(k_x) + M]^2}{[\mu_d(k_x) - M]^2+\xi_{SO}(k_x)^2+v^2}. \label{eq:rhoN_ratio_so_approx}
\end{align}
Let us first examine two limiting cases of Eq.~\eqref{eq:rhoN_ratio_so_approx}. If $\xi_{SO}(k_x)^2 \gg [\mu_d(k_x) - M]^2$, which implies $\mu_d(k_x) \approx M$, as well as $k_x\approx \pm k_0$ by Eq.~\eqref{eq:k0def_so}, then ${\rho_{\uparrow}^{(N)}(k_x, E=0)}/{\rho_{\downarrow}^{(N)}(k_x, E=0)} \approx {\rho_{\uparrow}^{(M)}}/{\rho_{\downarrow}^{(M)}}$; if $\max[\xi_{SO}(k_x)^2] \ll [\mu_d(k_x) - M]^2 \ll 4M^2$, then ${\rho_{\uparrow}^{(N)}(k_x, E=0)}/{\rho_{\downarrow}^{(N)}(k_x, E=0)} \approx \frac{4M^2}{[\mu_d(k_x) - M]^2+v^2} < {\rho_{\uparrow}^{(M)}}/{\rho_{\downarrow}^{(M)}}$. In general cases, an inequality similar to Eq.~\eqref{eq:R_ratio_ineq} holds for each $k_x$ with $v^2$ replaced by $\xi_{SO}(k_x)^2+v^2$ but a direct comparison between ${\rho_{\uparrow}^{(N)}(k_x, E=0)}/{\rho_{\downarrow}^{(N)}(k_x, E=0)}$ and ${\rho_{\uparrow}^{(M)}}/{\rho_{\downarrow}^{(M)}}$ is not available without making a specific assumption about the forms of $\xi_{SO}(k_x)$ and $\xi_{d}(k_x)$. Nonetheless, we expect that as long as $\xi_{d}(k_x)$ (hence $\mu_{d}(k_x)$) varies faster than $\xi_{SO}(k_x)$ with respect to $k_x$ around $\pm k_0$ -- which roughly requires the pristine $d$-orbital bandwidth to be larger than the spin-orbit coupling strength -- ${\rho_{\uparrow}^{(M)}}/{\rho_{\downarrow}^{(M)}}$ is approximately a maximum of ${\rho_{\uparrow}^{(N)}(k_x, E=0)}/{\rho_{\downarrow}^{(N)}(k_x, E=0)}$. Hence ${\rho_{\uparrow}^{(M)}}/{\rho_{\downarrow}^{(M)}} > {R_{\uparrow}}/{R_{\downarrow}}$ in the realistic range of spin-orbit coupling strength (see Fig.~\ref{fig:ratio_compare}). Therefore, by virtue of Eq.~\eqref{eq:dG}, we conclude 
\begin{align}
  \delta G(x\rightarrow 0, E=0) > 0\quad (\textit{Majorana zero mode}).
\end{align}
This is in sharp contrast to the vanishing $\delta G(E=0)$ in the single magnetic impurity case (cf. Eq.~\eqref{eq:deltaG_sym}), and is what we propose as a robust feature to distinguish a Majorana zero mode from trivial Shiba states accidentally occurring at zero energy by using the spin-polarized STM technique.

Furthermore, away from the end of the magnetic impurity chain, since the only $x$-dependence of $\delta G$ in Eq.~\eqref{eq:dG} comes from $\tilde{\rho}_{\uparrow/\downarrow}$ (with $R_{\uparrow/\downarrow}$ in Eq.~\eqref{eq:R_band_normal} independent on $x$), it is straightforward to see that, at zero-energy, the positive $\delta G$ decays exponentially from the end in the same way as the decay of the amplitude of the Majorana zero mode (see Eq.~\eqref{eq:rhoM_decay}). As a consequence, we expect a vanishing $\delta G(E=0)$ in the middle part of the chain that is sufficiently far ($>\lambda$ with $\lambda$ defined in Eq.~\eqref{eq:lambda_so}) from the end. Meanwhile, measurements in the middle of the chain at finite bias voltages ($0<|E|<\Delta$) will exhibit features associated with the Shiba-band spin densities. Our analysis detailed in Sec.~\ref{ssec:shiba_band} directly implies that these features will be dictated by the Van Hover singularities of the Shiba bands because of their dominant contributions to the DOS factors $\rho_{\text{DOS}}^{\uparrow/\downarrow}$ (see Eq.~\eqref{eq:rho_band}). The combination of the particle-hole symmetry and the sum rules generically leads to a sign reversal of $\tilde{\rho}_{\uparrow} - \tilde{\rho}_\downarrow$ across the induced $p$-wave gap (cf. Fig.~\ref{fig:rhos_band} (d)), and hence a sign reversal of $\delta G$ around zero energy, with peaked $\delta G$ of opposite signs attached to the Van Hove singularities at opposite energies. This is another robust feature that we expect in a spin-polarized STM measurement on the Shiba chain.

\section{Simulations}\label{sec:sim}

In this section we show numerical results that address several analytically intractable issues. These issues include a generic, realistic, parameter setting, the effect of a finite-size magnetic impurity chain, and the full energy and position dependence of the spin densities in the chain. our simulations also examine one special but important scenario which was not included in the previous analytic analysis. In this scenario, an impurity present at the end of the chain is strongly coupled to the chain but still artificially tuned to zero energy. This scenario involves fine-tuning, as levels usually repel, especially when strongly coupled to the atomic chain. It is also exponentially unlikely to happen in every chain in the experiment. The zero-energy Shiba states in this proposed scenario are apparently different from Shiba states induced by an isolated single magnetic impurity, which we have investigated analytically in the preceding sections, because of their strong coupling to the chain. This strong coupling to the chain changes: first, the level repulsion of every level in the impurity; second, the wavefunction of the impurity which is now strongly hybridized with the chain. We will use specific examples to demonstrate the quantitative yet significant difference between such spurious end states and the actual MZMs.

\begin{figure}
  \centering
  \includegraphics[width=0.3\textwidth]{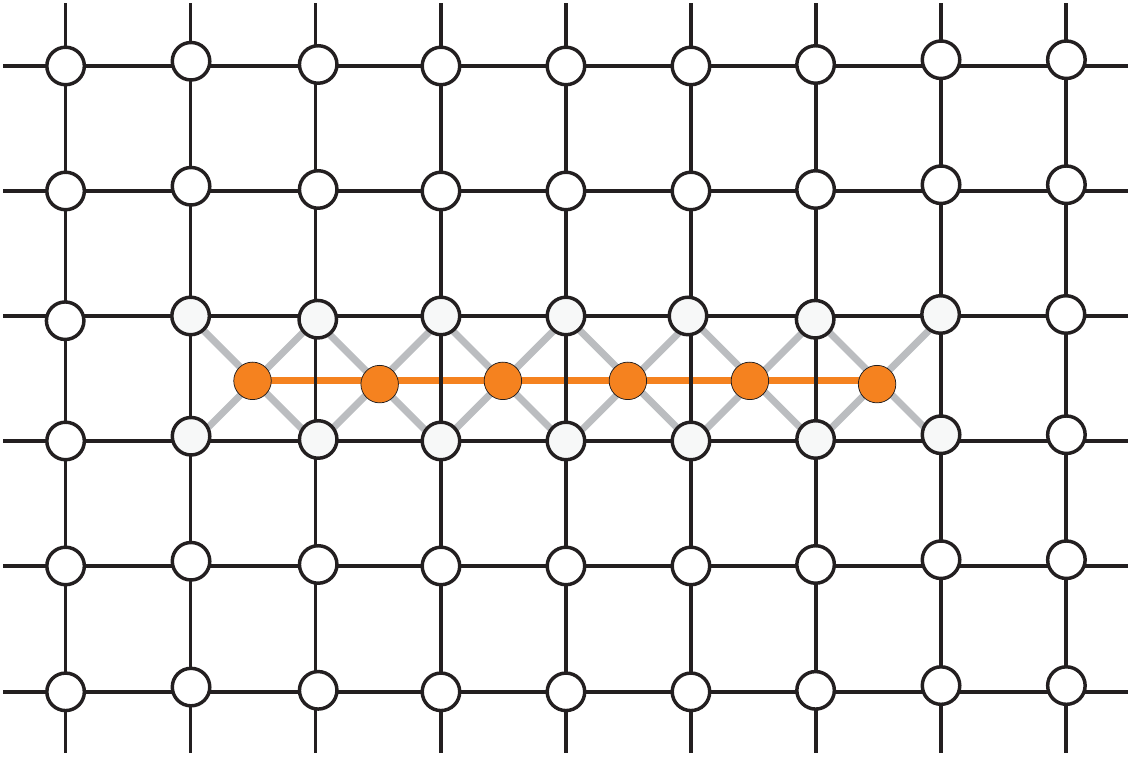}
  \caption{Schematic simulation setup. The red dots stand for magnetic impurity sites, and the white dots stand for superconductor sites. The superconductor is two-dimensional and infinite both dimensions; the magnetic impurities form a straight finite-size chain; the coupling between the two is only through nearest neighbors.}
  \label{fig:sim_setup}
\end{figure}

Our simulations are based on a geometry illustrated in Fig.~\ref{fig:sim_setup}, and the Hamiltonian is similar to a discretized version of Eqs.~\eqref{eq:ham_full}-\eqref{eq:ham_coup} with
\begin{align}
  &H_s(\bm{k}) = [2t_s (2 - \cos k_x - \cos k_y) - \mu_s \nn \\
  &\qquad\qquad + t_{SO} (\sin k_x \sigma_y - \sin k_y \sigma_x)]\otimes\tau_z+\Delta\tau_x, \label{eq:Hs_sim} \\
  &H_d(k_x) = M\sigma_z + (2t_d \cos k_x - \mu)\tau_z, \label{eq:Hd_sim} \\
  &\hat{H}_T = \sum_{<\bm{r}_s,\,\bm{r}_d>} V (\bm{c}_{\bm{r}_s}^\dag \bm{d}_{\bm{r}_d} - \bar{\bm{c}}_{\bm{r}_s}^\dag \bar{\bm{d}}_{\bm{r}_d}) + h.c., \label{eq:HT_sim}
\end{align}
where $\bm{r}_s$ and $\bm{r}_d$ are the positions of the superconductor and the magnetic impurity sites, respectively, and $<,>$ stands for nearest neighboring sites. Note that in the above Hamiltonian, both the spin-orbit coupling and the pairing potential are introduced only into the superconducting host. This is more realistic than our model used for analytical purposes (cf. Eqs.~\eqref{eq:Hs} and \eqref{eq:Hd}) but does not alter any physical consequences discussed previously. The Green's function of the hybrid system, with the superconducting host being infinite in two dimensions, is calculated numerically by using standard Dyson equations (see, e.g., Ref.~\onlinecite{li_topological_2014}). The spectral functions and spin densities are then obtained following the definitions Eq.~\eqref{eq:rho_def}. The spin densities thus obtained are exact in the current model, and have both energetic and spatial resolutions.

\begin{figure}
  \centering
  \includegraphics[width=0.5\textwidth]{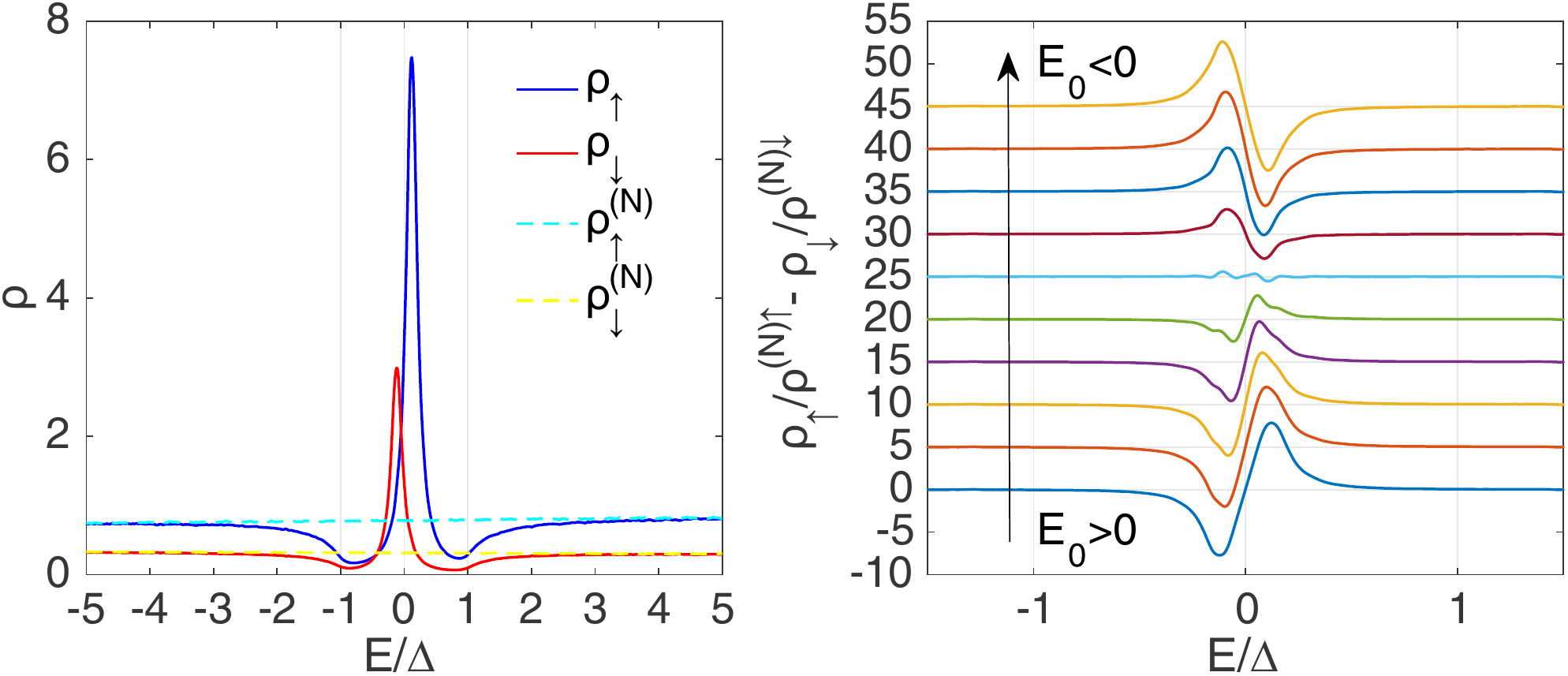}
  \caption{An example of the spin densities of a single magnetic impurity (the left panel), both in the superconducting (solid lines) and in the normal states (broken lines), and the difference of the normalized spin densities with various parameter settings (the right panel). Here, all spin densities are plotted as a function of energy, and the data in the right panel have been incrementally shifted by 5 for clear presentation. The parameter $V$ (see Eq.~\eqref{eq:HT_sim}) in the left panel is 0.6, and in the right panel $V$ increases from 0.6 to 0.69 with equal increments, resulting in a reversal of the sign of $E_0$ associated with the spin-$\uparrow$ peak. The parameters used in all plots (see Eqs.~\eqref{eq:Hs_sim} and \eqref{eq:Hd_sim}), except for $V$, are: $M=1$, $\mu=0.9$, $t_s=1$, $\mu_s=3$, $t_{SO} = 0.03$, $\Delta = 0.01$, and $\eta = 0.001$.}
  \label{fig:sim_single}
\end{figure}

To begin with, we check the case of a single magnetic impurity. The spin densities obtained in this case for one example parameter setting are shown in the left panel of Fig.~\ref{fig:sim_single}, where the generic features are fully consistent with our analytical solutions presented in Sec.~\ref{sec:single} and exemplified in Fig.~\ref{fig:rhos_single}. Namely, the in-gap spin densities appear as sharp peaks (delta functions in the limit of infinite lifetime and zero temperature) centered at opposite energies $\pm E_0$ for the two spins $\uparrow/\downarrow$, and the spin densities outside the superconducting gap converge to their normal-state values at energies sufficiently large compared with $\Delta$. The sum rules and their consequences in this case are most easily seen by plotting the difference between the normalized spin densities $\delta\tilde{\rho} \equiv \rho_{\uparrow}/\rho_{\uparrow}^{(N)} - \rho_{\downarrow}/\rho_{\downarrow}^{(N)}$, shown in the right panel of Fig.~\ref{fig:sim_single}, for different parameter settings that result in a crossover of $E_0$ from being positive to being negative. Clearly, $\delta\tilde{\rho}$ appears always as an antisymmetric function of energy, which is equivalent to Eq.~\eqref{eq:deltaG_sym}, and, in particular, $\delta\tilde{\rho}$ vanishes at zero energy.

\begin{figure}
  \centering
  \includegraphics[width=0.5\textwidth]{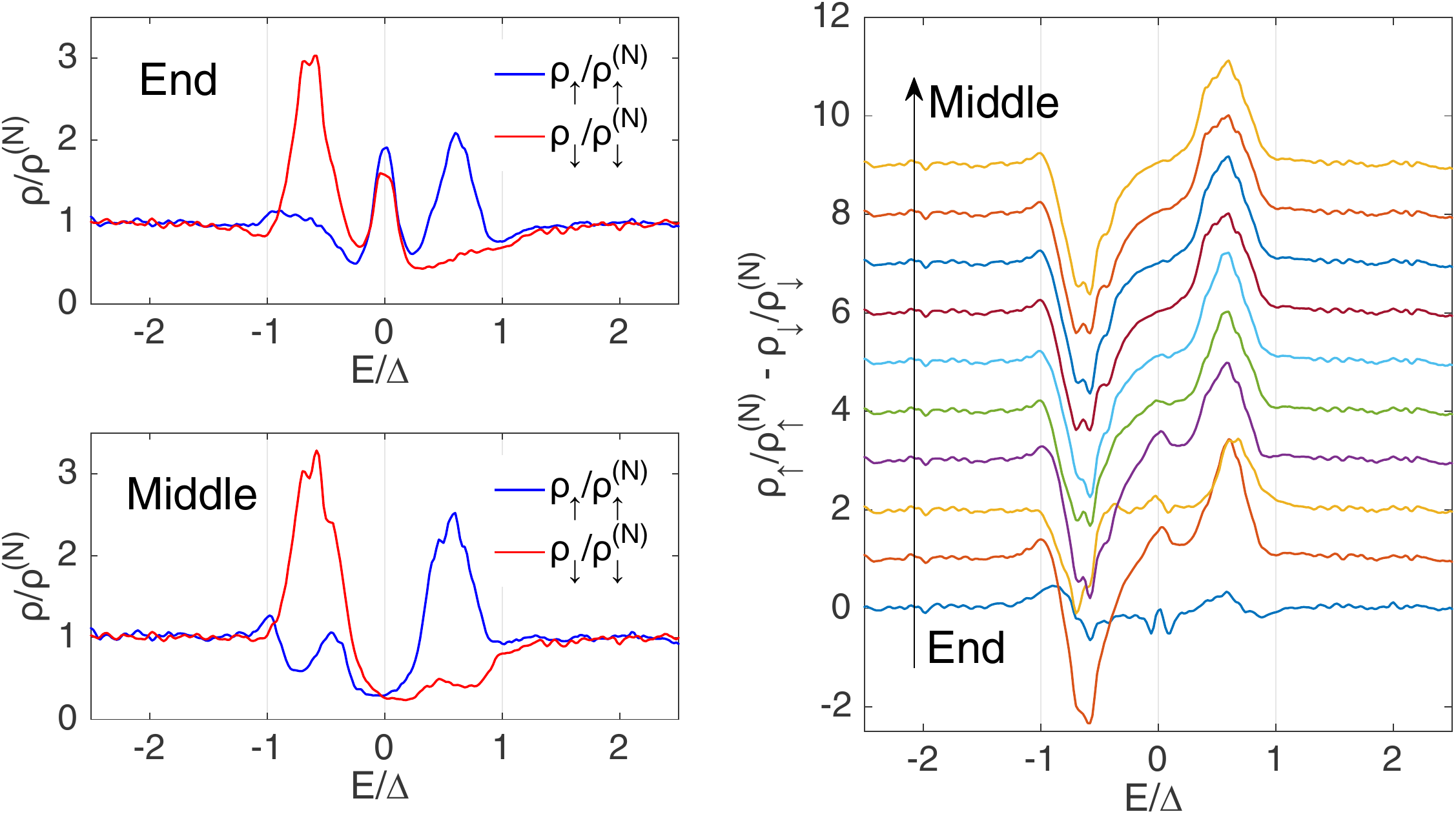}
  \caption{An example of the normalized spin densities at the end (4 sites averaged; the upper left panel) and in the middle (20 sites averaged; the lower left panel) of the magnetic impurity chain, and the difference of the normalized spin densities at different positions of the chain (showing the first 10 site of the chain; the right panel). Here, all spin densities are plotted as a function of energy, and the data in the right panel have been incrementally shifted by 1 for clear presentation. In this example, the chain is 60-site long in full, and the other parameters, corresponding to Eqs.~\eqref{eq:Hs_sim} to \eqref{eq:HT_sim}, are: $M=1$, $t_d=0.1$, $\mu=1.15$, $t_s=1$, $\mu_s=3$, $t_{SO} = 0.4$, $\Delta = 0.01$, $V=0.35$ and $\eta = 0.001$.}
  \label{fig:sim_chain}
\end{figure}

Next we simulate the case of a topologically nontrivial finite-length magnetic impurity chain. One such example is shown in Fig.~\ref{fig:sim_chain}. First we see in the two left panels of Fig.~\ref{fig:sim_chain} a comparison between the normalized spin densities at the end of the chain and those in the middle of the chain. The most obvious contrast is the presence of zero-energy peaks at the end but not in the middle, meanwhile the two spin densities exhibit a clear difference in their zero-energy peak values. Away from zero energy, the spin densities appear similar at the end and in the middle of the chain, with two dominant peaks associated with the Van Hove singularities of the Shiba bands -- these peaks are centered at opposite energies for opposite spins. Note that, as a consequence of the emergence of the localized MZM, the finite-energy Shiba-band peaks occur with reduced weights at the end compared with those in the middle of the chain.  In the right panel of Fig.~\ref{fig:sim_chain}, we further show the difference of the normalized spin densities $\delta\tilde{\rho}$, corresponding to the $\delta G$ in spin-polarized STM measurements (see Eq.~\eqref{eq:dG}), as a function of energy and the position in the chain. On the first few ($\sim 4$) sites of the chain, we clearly see a peak of $\delta\tilde{\rho}$ at zero energy, which oscillates fast due to strong spin-orbit coupling, and vanishes beyond about 6 sites. This zero-energy peak is pronounced in spite of the strong finite-energy peak (valley) contributed by the Shiba-band states -- the normal-state background spin densities $\rho_{\uparrow/\downarrow}^{(N)}$ are essentially energy independent, therefore the large DOS associated with the Van Hove singularities of the Shiba bands remains a dominant factor [cf. Eq.~\eqref{eq:rho_band}] in the normalized spin densities $\rho_{\uparrow/\downarrow}(E)/\rho_{\uparrow/\downarrow}^{(N)}$. The robust peak in $\delta\tilde{\rho}(E=0)$ sharply contrasts MZMs with Shiba states induced by single magnetic impurities, since the latter always leads to a vanishing $\delta\tilde{\rho}(E=0)$ owing to the sum rule.

\begin{figure}
  \centering
  \includegraphics[width=0.5\textwidth]{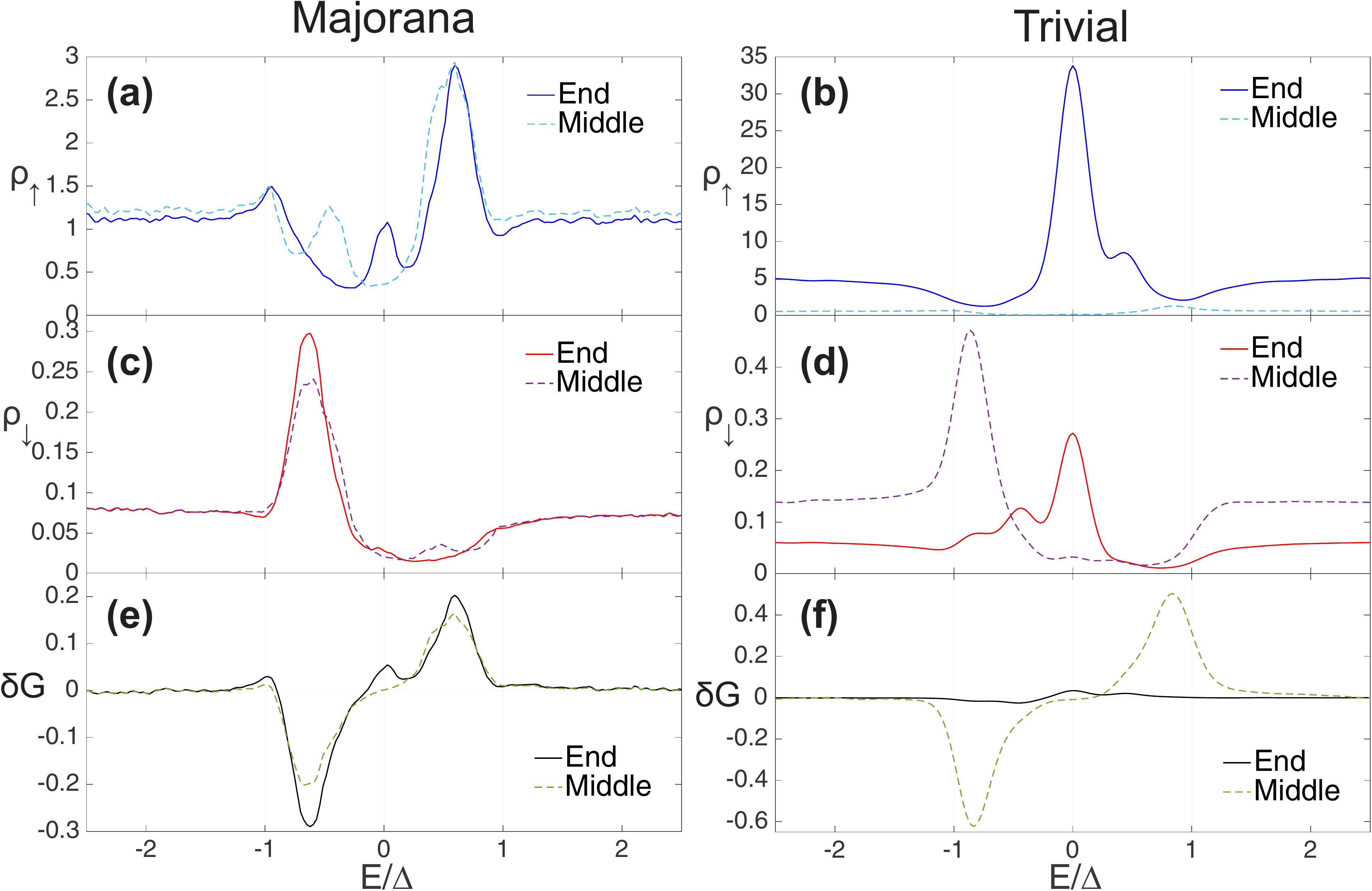}
  \caption{Comparison between MZMs (a, c and e) and artificial zero-energy end states (b, d and f) in terms of the spin densities [spin-$\uparrow$ in (a) and (b), and spin-$\downarrow$ in (c) and (d)] and $\delta{G}$ [see  Eq.~\eqref{eq:dG}] that can be extracted from spin-polarized STM measurements. In each plot we show simulation data for both the end (solid lines) and the middle (broken lines) of the magnetic impurity chain. The parameters used for the nontrivial chain are exactly the same as those used in Fig.~\ref{fig:sim_chain}, and we have taken the second site [see the right panel of Fig.~\ref{fig:sim_chain}, as well as Eqs.~\eqref{eq:wf_maj_so}-\eqref{eq:lambda_so}] as the end of the chain. The parameters used for the trivial chain, which is also 60-site long, are: $M=1$, $t_d=0.1$, $\mu=0.5$, $t_s=1$, $\mu_s=3$, $t_{SO} = 0.1$, $\Delta = 0.002$, $V=0.3$ and $\eta = 0.0001$. The local potential on the first site of the trivial chain that induces zero-energy Shiba states is found to be $-0.51$. The additional parameters used in obtaining $\delta{G}$ are $\tilde{w}_{N} = \tilde{w}_{P}^{-1} = 2$.}
  \label{fig:sim_comp}
\end{figure}

Now we address another possible but unlikely scenario of trivial zero-energy end states. Namely, we simulate the case of Shiba states tuned to zero energy by a local potential [nonzero only at the first site of the chain and proportional to $\sigma_0\otimes\tau_z$ as in Eq.~\eqref{eq:Hd_sim}], but meanwhile strongly coupled to the rest of a topologically trivial chain. This case is different from the case of an isolated single magnetic impurity, which has been discussed in the preceding part of the paper, because the coupling between the local impurity states and the extended states in the chain leads to local spin densities necessarily containing both contributions. Our sum rules (Eqs.~\eqref{eq:sum_rule0} and \eqref{eq:sum_rule_full}) established in our previous discussion which considered either totally local or totally extended Shiba states become invalid. Apparently this creates an artificial zero-energy end state that is indistinguishable from a genuine MZM even in terms of the spin signature proposed in this paper, because the background normal-state spin densities are now extended (unlike the local Shiba state case) and cannot screen the local Shiba ones, unlike in Eq.~\eqref{eq:normalized_rho}. We show, however, that this expectation is not true by using typical simulation results. 

Before presenting the results we first discuss a general aspect of our simulation. Our artificial zero-energy end states require simultaneously three conditions: a trivial chain (cf. Eq.~\eqref{eq:topo_cond_main}) which does not host any localized zero-energy state by itself; an impurity at the end of the chain that is strongly coupled to the chain; a local potential at the site of the end impurity that is strong enough to induce zero-energy Shiba states. Here, the combination of the first two conditions works in general against the third condition, and the localized end states need fine detuning parameters in order to occur at zero energy. When such zero-energy end states do occur, as we show in the right panels of Fig.~\ref{fig:sim_comp} with one example, the spin densities at the end are dominated by the local impurity states. More specifically, the density of one spin (assumed to be spin-$\uparrow$ as in the example in Fig.~\ref{fig:sim_comp}) has a much larger magnitude than the other spin, or the bulk states in the chain, not only in the subgap regime but in the energy range comparable to the coupling energy between the impurity and the superconducting host (see Fig.~\ref{fig:sim_comp} (b) and (d); note the scale of the $y$-axis in each plot). This is because that, roughly speaking, the occurrence of zero-energy Shiba states is always a consequence of the chemical potential becoming sufficiently close to one of the spin-polarized pristine energy levels, as can be seen from Eq.~\eqref{eq:E0_shiba_main}. The local dominance of spin-$\uparrow$ density leads to, at the end of the chain, a strongly enhanced overall low-energy spectral density which is, by definition, the sum of the spin densities; however it also leads to a much larger local ratio $R_{\uparrow}/R_{\downarrow}$, appearing in the denominator of $\delta{G}$ in Eq.~\eqref{eq:dG}, than the case of MZMs where $R_{\uparrow/\downarrow}$ are determined by the $d$-orbital bands [see Eq.~\eqref{eq:R_band_normal}]. This in turn heavily suppresses $\delta{G}$ at the end of chain, especially when compared with $\delta{G}$ in the middle of the chain (see Fig.~\ref{fig:sim_comp}(f)), despite of the fact that the normalized spin densities $\tilde{\rho}_{\uparrow/\downarrow}$ do not cancel each other due to the absence of a sum rule in this case for the subgap states close to the end. In contrast to the artificial zero-energy end states, in the left panels of Fig.~\ref{fig:sim_comp} the spin densities and $\delta{G}$ obtained in a topologically nontrivial chain (with the same parameters as in the example in Fig.~\ref{fig:sim_chain}) exhibit similar magnitudes at the end and in the middle of the chain, but differs crucially by the presence of a pronounced zero-energy peak.

\section{Conclusion}

In summary, we have systematically investigated the spin properties of Shiba states and Majorana zero modes associated with quantum magnetic impurities/adatoms embedded in a conventional superconductor. In particular we have formulated the sum rules that relate the spin densities in the superconducting states to those in the normal states; we then used these relations to understand the outcomes of spin-polarized scanning tunneling microscope measurements in the sequential electron tunneling regime. Based on this understanding, we propose a robust and definite spin signature that provides crucial test for distinguishing Majorana zero modes from trivial Shiba states accidentally tuned to zero energy.

\vspace{5mm}
\textit{Note added.} Recently, a spin-polarized STM study of chains of the transition metal cobalt (Co) on Pb(110) was reported by Ruby \textit{et al.} \cite{ruby_exploring_2017}, where no MZM has been observed because the band structure of the Co chains leads to a topologically trivial superconducting phase. Nevertheless, the observed spin contrast features associated with the Shiba bands in this experiment are consistent with our results here. Upon finishing this work, we have also noticed the appearance of two other preprints proposing spin-polarized STM measurement in distinguishing MZMs and trivial in-gap quasiparticle states \cite{devillard_majorana_2017, maska_spin-polarized_2017}. Both of these two preprints concern the Andreev reflection regime, which is different from the single-electron sequential tunneling regime focused on in this paper -- the experiment reported in Ref.~\onlinecite{Jeong_2017_to_appear} was performed in the latter regime.

\acknowledgements

This work has been supported by ONR-N00014-14-1-0330, ONR-N00014-11-1-0635, ONR- N00014-13-10661, NSF-MRSEC programs through the Princeton Center for Complex Materials DMR-1420541, NSF-DMR-1608848, Department of Energy de-sc0016239, Simons Investigator Award, NSF EAGER Award NOA - AWD1004957, DOE-BES, Packard Foundation, Schmidt Fund for Innovative Research, ARO-MURI program W911NF-12-1-046, Gordon and Betty Moore Foundation as part of EPiQS initiative (GBMF4530), and Eric and Wendy Schmidt Transformative Technology Fund at Princeton. BAB wishes to thank Ecole Normale Superieure, UPMC Paris, and Donostia International Physics Center for their generous sabbatical hosting.

\appendix

\section{Derivations of the Green's functions of the magnetic impurity and the uncoupled superconductor}\label{app:GF}
We first derive Eq.~\eqref{eq:Gd0} for the Green's function of the magnetic impurity (the $d$-orbital degrees of freedom) described by the general Hamiltonian \eqref{eq:ham_full0} with the coupling term \eqref{eq:ham_coup0}. In the Nambu basis, the retarded Green's function in the hybrid system is defined as
\begin{align}
  &\hspace{-2mm} G_{\psi'\psi^\dag}(E^+) = \int\limits_{-\infty}^{+\infty} dt\; e^{i E^+ t} \left[-i \theta(t)\langle\{\psi'(t), \psi^\dag(0)\}\rangle\right],
\end{align}
where each of $\psi$ and $\psi'$ can be any component of $\bm{c}$, $\bar{\bm{c}}$, $\bm{d}$ or $\bar{\bm{d}}$. We shall denote the matrix form of $G_{\psi'\psi^\dag}(E^+)$ by $G_d(E^+)$ when $\psi$ and $\psi'$ are constrained to the components of $\bm{d}$ or $\bar{\bm{d}}$, by $G_{sd}(E^+; \bm{r})$ when $\psi$ is constrained to the components of $\bm{d}$ or $\bar{\bm{d}}$ and $\psi'$ is constrained to the components of $\bm{c}_{\bm{r}}$ or $\bar{\bm{c}}_{\bm{r}}$,  and by $G_{s}(E^+; \bm{r}', \bm{r})$ when $\psi$ ($\psi'$) is constrained to the components of $\bm{c}_{\bm{r}}$ or $\bar{\bm{c}}_{\bm{r}}$ ($\bm{c}_{\bm{r}'}$ or $\bar{\bm{c}}_{\bm{r}'}$). We will also denote the Green's functions in the decoupled limit, namely, $V=0$ in Eq.~\eqref{eq:ham_coup0}, by a superscript $(0)$. With these notations, the Dyson equations for the hybrid system reads
\begin{align}
  &G_d(E^+) \nn \\
  &= G_d^{(0)}(E^+) +\int d\bm{r}\; G_d^{(0)}(E^+)[V\delta(\bm{r})]G_{sd}(E^+;\bm{r}) \\
  &= G_d^{(0)}(E^+) + V G_d^{(0)}(E^+) G_{sd}(E^+;\bm{r}=0), \\
  &G_{sd}(E^+;\bm{r}) \nn \\
  &= \int d\bm{r}'\; G_s^{(0)}(E^+; \bm{r}, \bm{r}')[V\delta(\bm{r}')]G_{d}(E^+) \\
  &= V G_s^{(0)}(E^+; \bm{r}) G_{d}(E^+),
\end{align}
therefore
\begin{align}
  G_d(E^+) &= G_d^{(0)}(E^+) \nn \\
  &+ V^2 G_d^{(0)}(E^+) G_s^{(0)}(E^+; \bm{r}=0) G_{d}(E^+).
\end{align}
In the above equations, we have used $G_{sd}^{(0)}(E^+;\bm{r}) = 0$, and we have shortened the notation $G_s^{(0)}(E^+; \bm{r}, 0)$ to $G_s^{(0)}(E^+; \bm{r})$ by using the fact that in the decoupled limit the superconductor is translational invariant (we assume that the superconductor has no surfaces). Noticing that
\begin{align}
  G_d^{(0)}(E^+) = (E^+ - H_d)^{-1},
\end{align}
we obtain
\begin{align}
  G_d(E^+) = \left[E^+-H_d-V^2G_s^{(0)}(E^+; \bm{r}=0)\right]^{-1}. \label{eq:Gd00}
\end{align}

The Green's function $G_s^{(0)}(E^+; \bm{r})$ for a three-dimensional superconductor has been derived by Pientka et al. \cite{pientka_topological_2013}. Here we derive $G_s^{(0)}(E^+; \bm{r})$ for a two-dimensional superconductor as follows. By definition,
\begin{align}
  &G_s^{(0)}(E^+;\bm{r}) = \int d\bm{k} \; G_s^{(0)}(E^+;\bm{k}) e^{i\bm{k}\cdot\bm{r}} \\
  =& \int_0^{\infty} \frac{k\,dk}{2\pi} \int_0^{2\pi} \frac{d\varphi}{2\pi} \; [E^+-H_s(k)]^{-1}  e^{i k r \cos\varphi}.
\end{align}
To perform the above integral, we will make use of the following formulas:
\begin{align}
  &\int_0^{\infty} \frac{k\,dk}{2\pi} e^{i k x} [E^+-\xi(k)\tau_z-\Delta\tau_x]^{-1} \nonumber \\
  &\qquad\simeq -\rho \int d\xi \, e^{i (k_F+\xi/v_F) x} \,\frac{E^++\Delta\tau_x+\xi\tau_z}{\xi^2+[\Delta^2-(E^+)^2]}, \\
 &\int d\xi \, e^{i \xi x/v_F} \,\frac{1}{\xi^2+a^2} = \frac{\pi}{a}e^{-a|x|/v_F},\quad \text{Re}(a)>0, \\
 &\int d\xi \, e^{i \xi x/v_F} \,\frac{\xi}{\xi^2+a^2} \frac{\omega_c^2}{\xi^2+\omega_c^2} \quad (\text{Re}(a)>0)\nonumber \\
 &\qquad = \frac{i\pi \omega_c^2}{\omega_c^2-a^2}\,\text{sgn}(x)(e^{-a|x|/v_F}-e^{-\omega_c|x|/v_F}), \\
 &I(z) \equiv \int_{-\pi/2}^{\pi/2} d\varphi \; e^{i z \cos\varphi} \nn \\
 &\quad = \pi J_{0}(z) + 2i\sum_{m=-\infty}^{+\infty} \frac{J_{2m+1}(z)}{2m+1},
\end{align}
where $x = r\cos\varphi$, $\xi(k)$ is the normal-state dispersion relation, $k_F$ is the Fermi wave vector defined by $\xi(k_F)=0$, $v_F=(\partial \xi/\partial k)|_{k=k_F}$ is the Fermi velocity, $\rho = k_F/2\pi v_F$ is the density of states at the Fermi energy, $\omega_c$ is a cut-off frequency which will be sent to $+\infty$ in the end, and $J_n(z)$ is the the $n$-th order Bessel function of the first kind.

If we assume $H_s(k)$ to be given by Eq.~\eqref{eq:Hs}, then $\xi(k) = t_s k^2-\mu_s$, and $k_{F} = \sqrt{\mu_s/t_s}$, $v_{F} = 2t_s k_{F}$, $\rho=1/4\pi t_s$. We obtain
\begin{align}
  G_s^{(0)}(E^+,k_x;\bm{r}) = g_0\tau_0+g_x\tau_x+g_z\tau_z,
\end{align}
where
\begin{align}
  &g_{0} = -\frac{\rho}{2}\,\frac{E^+}{\xi_E}[I(k_{F}r+i\xi_Er/v_{F})+I(-k_{F}r+i\xi_Er/v_{F})], \\
  &g_{x} = -\frac{\rho}{2}\,\frac{\Delta}{\xi_E}[I(k_{F}r+i\xi_Er/v_{F})+I(-k_{F}r+i\xi_Er/v_{F})], \\
  &g_{z} = - \frac{\rho}{2} \, \frac{i\omega_c^2}{\omega_c^2-\xi_E^2}
  \Bigl[ I(k_{F}r+i\xi_Er/v_{F}) - I(-k_{F}r+i\xi_Er/v_{F}) \nonumber \\
  &\qquad- I(k_{F}r+i\omega_c r/v_{F}) + I(-k_{F}r+i\omega_c r/v_{F}) \Bigr], \\
  &\xi_E = \sqrt{\Delta^2-(E^+)^2}, \quad \text{Re}(\xi_E)>0.
\end{align}
Here, $E$ is not limited to the subgap energy range, but can be arbitrary instead.

Particularly, when $r\rightarrow0$,
\begin{align}
  G_s^{(0)}(E^+;\bm{r}=0) =  -\frac{\pi\rho}{\xi_E}(E^+\tau_0+\Delta\tau_x), \label{eq:Gs0}
\end{align}
which has the same form as $G_s^{(0)}(E^+;\bm{r}=0)$ in the 3D case. By substituting Eq.~\eqref{eq:Gs0} into Eq.~\eqref{eq:Gd00} and changing the notation $\rho$ to $\rho_s$, we obtain Eqs.~\eqref{eq:Gd0} and \eqref{eq:v0}.

\section{solutions of the single magnetic impurity model}\label{app:single}

By using the explicit form of $H_d$ in Eq.~\eqref{eq:Hd0}, we can decompose Eq.~\eqref{eq:Gd0} into a block-diagonal form
\begin{align}
  &G_d(E^+) = g_{d-}(E^+) \oplus g_{d+}(E^+), \\
  &g_{d\mp}(E^+) = \left[(E^+ \mp M)\tau_0 + \mu\tau_z + v\,\frac{E^+\tau_0+\Delta\tau_x}{\sqrt{\Delta^2-(E^+)^2}}\right]^{-1} \nn\\
  &\;=\frac{1}{D_{\mp}}\left[(E^+ \mp M)\tau_0 - \mu\tau_z + v\,\frac{E^+\tau_0-\Delta\tau_x}{\sqrt{\Delta^2-(E^+)^2}}\right], \\
  &D_{\mp} = (E^+ \mp M)^2 - \mu^2 - v^2 + \frac{2vE^+(E^+ \mp M)}{\sqrt{\Delta^2-(E^+)^2}}.
\end{align}
Here, $g_{d\mp}$ correspond to the components $(d_{\uparrow}^\dag, d_{\downarrow})$ and $(d_{\downarrow}^\dag, -d_{\uparrow})$, respectively. Thus, by definition Eq.~\eqref{eq:rho_def}, we have
\begin{align}\label{eq:rho_general}
  \rho_{\uparrow/\downarrow}(E) = -\frac{1}{\pi}\text{Im}\bigl[(E^+ \mp M - \mu + v_{E^+})/D_{\mp}\bigr]_{\eta\rightarrow 0}.
\end{align}
where $v_{E^+} = v E^+/\sqrt{\Delta^2-(E^+)^2}$. The above expression can be simplified in two energy ranges separately.

If $|E|>\Delta$, then $E^+/\sqrt{\Delta^2-(E^+)^2} = i |E|/\sqrt{E^2-\Delta^2} + O(\eta)$, it follows that
\begin{align}\label{eq:rho_supergap}
  \rho_{\uparrow/\downarrow}(|E|>\Delta) = \frac{|v_E| [(E \mp M - \mu)^2 + v^2] / \pi}{[(E \mp M)^2 - \mu^2 - v^2]^2 + 4 |v_E|^2 (E \mp M)^2},
\end{align}
where $|v_E| = v/\sqrt{1-\Delta^2/E^2}$. Particularly, if $|E|\rightarrow\Delta$, we find
\begin{align}
   \rho_{\uparrow/\downarrow}(|E|=\Delta_+) \simeq \frac{(\mu \pm M)^2 + v^2}{2\sqrt{2} \pi v M^2} \sqrt{|E|/\Delta-1},
\end{align}
where we have used the assumption $M, v \gg \Delta$; if $|E|\gg\Delta$, such that $|v_E| \simeq v$, we find
\begin{align}
   \rho_{\uparrow/\downarrow}(|E|\gg\Delta) \simeq \frac{v / \pi}{(E \mp M + \mu)^2 + v^2},
\end{align}
which are the broadened spin densities with a Lorentzian function of width $v$ as in a normal state.

If $|E|<\Delta$, then the Green's function for the pristine superconductor contains no poles, hence we may drop the infinitesimal imaginary part in $E^+/\sqrt{\Delta^2-(E^+)^2}$ and substitute $v_{E^+}$ by $v_E$, and Eq.~\eqref{eq:rho_general} can be rewritten as
\begin{align}\label{eq:rho_subgap_decomp}
  \rho_{\uparrow/\downarrow}(E) &= -\frac{1}{2\pi}\text{Im}\Bigl(\frac{1 - \mu/\sqrt{\mu^2 + v^2 + v_E^2}}{E^+ \mp M +v_E - \sqrt{\mu^2 + v^2 + v_E^2}} \nn\\
  &\;+ \frac{1 + \mu/\sqrt{\mu^2 + v^2 + v_E^2}}{E^+ \mp M +v_E + \sqrt{\mu^2 + v^2 + v_E^2}}\Bigr)_{\eta\rightarrow 0}.
\end{align}
Since we are considering $|E| < \Delta \ll M$, we have $E - M +v_E - \sqrt{\mu^2 + v^2 + v_E^2}<0$ and $E + M +v_E + \sqrt{\mu^2 + v^2 + v_E^2}>0$, therefore Eq.~\eqref{eq:rho_subgap_decomp} becomes
\begin{align}
  &\rho_{\uparrow/\downarrow}(|E|<\Delta) = \frac{1}{2}(1 \pm \mu/\sqrt{\mu^2 + v^2 + v_E^2}) \nn\\
  &\qquad\qquad\cdot\delta(E \mp M +v_E \pm \sqrt{\mu^2 + v^2 + v_E^2}), \\
  &\quad\simeq \frac{v}{(\mu \mp M)^2 + v^2} \sqrt{\Delta^2 - E_0^2} \;\delta(E \mp E_0), \label{eq:rho_subgap}\\
  &E_0 \simeq \Delta\,\frac{M^2-\mu^2-v^2}{\sqrt{(M^2-\mu^2-v^2)^2+4M^2v^2}}, \label{eq:E0_shiba}
\end{align}
where we have used the approximation $E+v_E = E(1+v /\sqrt{\Delta^2-E^2}) \simeq v_E$. In the above equations, $\pm E_0$ is the energy of the Shiba states. In the case of $\mu=0$, Eq.~\eqref{eq:E0_shiba} becomes $E_0=-\Delta[1-(M/v)^2]/[1+(M/v)^2]$, which corresponds to the original solution of Yu \cite{yu_bound_1965}, Shiba \cite{shiba_classical_1968} and Rusinov \cite{rusinov_theory_1969}.

When $|E|=\Delta$, it is easy to verify directly from Eq.~\eqref{eq:rho_general} that
\begin{align}
  \rho_{\uparrow/\downarrow}(|E|=\Delta) = 0,
\end{align}
which is consistent with both Eq.~\eqref{eq:rho_supergap} and Eq.~\eqref{eq:rho_subgap} in the $|E|=\Delta$ limit.

Now we show the sum rule presented in Eq.~\eqref{eq:sum_rule_full}. The integrals of $\rho_{\uparrow/\downarrow}$ in the subgap energy range are straightforward and are given by Eq.~\eqref{eq:sum_rule0}. In the energy range above the gap, with the assumption $\Delta\le |E|\le E_c \ll v,M$, we approximate Eq.~\eqref{eq:rho_supergap} by
\begin{align}
  \rho_{\uparrow/\downarrow}(E) \simeq \frac{|v_E| [(\mu \pm M)^2 + v^2] / \pi}{(M^2 - \mu^2 - v^2)^2 + 4 |v_E|^2 M^2},
\end{align}
which is even in $E$. Thus
\begin{align}
  &\left(\int_{-E_c}^{-\Delta}+\int_{\Delta}^{E_c}\right)\rho_{\uparrow/\downarrow}(E)\,dE \simeq 2\int_{\Delta}^{E_c}\rho_{\uparrow/\downarrow}(E)\,dE \nn\\
  &\quad\simeq 2\int_{\Delta}^{E_c}dE\, \frac{|v_E| [(\mu \pm M)^2 + v^2] / \pi}{(M^2 - \mu^2 - v^2)^2 + 4 |v_E|^2 M^2} \nn\\
  &\quad= 2\Delta\int_{\Delta}^{E_c}dx\, \frac{v[(\mu \pm M)^2 + v^2] / \pi}{(M^2 - \mu^2 - v^2)^2 + 4 M^2 v^2 (1+\frac{1}{x^2})} \nn\\
  &\hspace{0.25\textwidth} (x\equiv \sqrt{\Bigl({\frac{E}{\Delta}}\Bigr)^2-1}) \nn\\
  &\quad=\frac{2\Delta v[(\mu \pm M)^2 + v^2] / \pi}{(M^2 - \mu^2 - v^2)^2 + 4 M^2 v^2} \int_{0}^{x_c}dx\, \frac{x^2}{x^2+a^2} \nn\\
  &\quad (x_c \equiv \sqrt{\Bigl(\frac{E_c}{\Delta}\Bigr)^2-1}, \; a\equiv \sqrt{{\frac{4 M^2 v^2}{(M^2 - \mu^2 - v^2)^2 + 4 M^2 v^2}}}) \nn\\
  &\quad=2\Delta\rho_{\uparrow/\downarrow}^{(N)}(E=0) \,(x_c - a\arctan{\frac{x_c}{a}}).
\end{align}
Assuming $E_c\gg\Delta$, and noticing $a=\sqrt{1-(\frac{E_0}{\Delta})^2}$ with $E_0$ given by Eq.~\eqref{eq:E0_shiba}, we have $x_c\simeq E_c/\Delta$ and $\arctan{\frac{x_c}{a}}\simeq \pi/2$, therefore
\begin{align}
  &\left(\int_{-E_c}^{-\Delta}+\int_{\Delta}^{E_c}\right)\rho_{\uparrow/\downarrow}(E)\,dE \nn\\
  &\qquad\simeq
  \rho_{\uparrow/\downarrow}^{(N)}(E=0) \,\Bigl(2 E_c - \pi \sqrt{\Delta^2-E_0^2}\Bigr).
\end{align}
Combining this equation with Eq.~\eqref{eq:sum_rule0}, we obtain Eq.~\eqref{eq:sum_rule_full}.

We now show that if $|M-\sqrt{\mu^2+v^2}|\ll v$, Eqs.~\eqref{eq:E0_shiba} and \eqref{eq:rho_subgap} (i.e. Eqs.~\eqref{eq:E0_shiba_main} and \eqref{eq:rho_subgap_main} in the main text) reduce to Eqs.~\eqref{eq:E_toy} and \eqref{eq:rho_toy} in the main text, respectively, with $\Delta_d$ replaced by $v$ and an energy scaling factor $\Delta/(\Delta+v) \simeq \Delta/v$ in Eq.~\eqref{eq:ham_deep_shiba}. First let us write $M-\sqrt{\mu^2+v^2} = \epsilon v$ with $|\epsilon|\ll 1$, then to the linear order in $\epsilon$, Eqs.~\eqref{eq:E0_shiba} becomes
\begin{align}
  E_0 \simeq \Delta\,\frac{2\epsilon M v}{2Mv} = \Delta\epsilon = \frac{\Delta}{v}(M-\sqrt{\mu^2+v^2}),
\end{align}
which is equivalent to Eq.~\eqref{eq:E_toy}. Then plugging the above expression into Eq.~\eqref{eq:rho_subgap}, we have
\begin{align}
  &\rho_{\uparrow/\downarrow}(E) \simeq \frac{v\sqrt{\Delta^2 (1-\epsilon^2)}}{(\mu \mp M)^2 + v^2} \;\delta(E - \frac{\Delta}{v}E_\pm).
\end{align}
By defining $E_d = (v/\Delta)E$, and keeping only the zeroth order terms in $\epsilon$, we further have
\begin{align}
  &\rho_{\uparrow/\downarrow}(E_d) \simeq \frac{\sqrt{\mu^2+v^2} \pm \mu}{2\sqrt{\mu^2+v^2}} \;\delta(E_d - E_\pm),
\end{align}
which is equivalent to Eq.~\eqref{eq:rho_toy}.

\section{Solutions of Majorana zero modes with perturbative spin-orbit coupling}\label{app:maj}

We solve the effective Hamiltonian in Eq.~\eqref{eq:ham_deep_shiba} in the vicinity of $\pm k_0$, where the superconducting gap is opened, to obtain the $d$-orbital components of the Majorana zero modes with perturbative spin-orbit coupling. By definition, $k_0$ satisfies Eq.~\eqref{eq:k0def}. It follows from the effective Hamiltonian Eq.~\eqref{eq:ham_deep_shiba} (which is indeed valid in the vicinity of $\pm k_0$), as well as the solutions of the toy model presented in Sec.~\ref{sec:toy}, that the $d$-orbital components of the eigenstates and their associated group velocities at $k_x = \pm k_0$, in the limit of vanishing spin-orbit coupling, are given by
\begin{align}
  &\psi_{+,\pm k_0}(x) =
  \begin{pmatrix}
    \cos\frac{\theta_0}{2} \\ 0 \\ \sin\frac{\theta_0}{2} \\ 0
  \end{pmatrix} e^{\pm i k_0 x},
  &\mathrm{v}_{+,\pm k_0} = \pm \mathrm{v}_0, \label{eq:psi+k0}\\
  &\psi_{-,\pm k_0}(x) =
  \begin{pmatrix}
    0 \\ \sin\frac{\theta_0}{2} \\ 0 \\ -\cos\frac{\theta_0}{2}
  \end{pmatrix} e^{\pm i k_0 x},
  &\mathrm{v}_{-,\pm k_0} = \mp \mathrm{v}_0, \label{eq:psi-k0}
\end{align}
where
\begin{align}
  &\sin\theta_0 = v/M,\quad \cos\theta_0 = \sqrt{M^2-v^2}/M, \\
  &\mathrm{v}_0 = (\partial E_0/\partial k_x)|_{k_x=k_0} = \frac{\Delta\cos\theta_0}{\Delta+v}  \mathrm{v}_d(k_0),
\end{align}
with $\mathrm{v}_d(k_0) = (\partial \xi_d/\partial k_x)|_{k_x=k_0}$ the bare group velocity of the pristine $d$-orbital band (in the absence of spin-orbit coupling) at $k_0$.

Based on the above solutions, we may choose a new basis for the low energy states in the chain to be $(\psi_{+, +k_x},\psi_{+, -k_x},\psi_{-, +k_x},\psi_{-, -k_x})$, where $\psi_{\pm, \pm k_x}$ are defined by replacing $k_0$ in $\psi_{\pm, \pm k_0}$ in Eqs.~\eqref{eq:psi+k0} and \eqref{eq:psi-k0} with $k_x$ that is in a (sufficiently small) neighborhood of $k_0$. The original spin-orbit coupling in Eq.~\eqref{eq:Hd}, rescaled by a factor of $\Delta/(\Delta+v)$ according to Eq.~\eqref{eq:ham_deep_shiba}, is then included perturbatively by a projection to this new basis and keeping only the leading order terms. Thus we obtain the following effective Hamiltonian for a nontrivial $p$-wave superconductor
\begin{align}
  &H_{\text{psc}} = \nn\\
  &\begin{pmatrix}
    \mathrm{v}_0(k_x - k_0) & 0 & -i\Delta_p & 0 \\
    0 & -\mathrm{v}_0(k_x + k_0) & 0 & i\Delta_p \\
    i\Delta_p & 0 & -\mathrm{v}_0(k_x - k_0) & 0 \\
    0 & -i\Delta_p & 0 & \mathrm{v}_0(k_x + k_0) \\
  \end{pmatrix}, \label{eq:Hpsc} \\
  &\Delta_p = \xi_{SO}(k_0)v\Delta/M(\Delta+v) \simeq \xi_{SO}(k_0)\Delta/M, \label{eq:Delta_p_pert}
\end{align}
where we have used the fact that $\xi_{SO}(k_x)$ is an odd function of $k_x$.

A Majorana zero mode corresponds to a zero-energy solution of the Hamiltonian~\eqref{eq:Hpsc} in real space with the boundary condition $\chi(x=0) = 0$ in the original basis. Such solutions can be easily obtained in the basis of Hamiltonian~\eqref{eq:Hpsc} to be (up to a normalization factor)
\begin{align}
  &\tilde{\chi}_1(x) =
  \begin{pmatrix}
    e^{ik_0x}\\ -e^{-ik_0x} \\ e^{ik_0x} \\ -e^{-ik_0x}
  \end{pmatrix} e^{- x/\lambda},\;
  \tilde{\chi}_2(x) =
  \begin{pmatrix}
    e^{ik_0x}\\ -e^{-ik_0x} \\ -e^{ik_0x} \\ e^{-ik_0x}
  \end{pmatrix} e^{x/\lambda}, \\
  &\lambda = \frac{\mathrm{v}_0}{\Delta_p} = \frac{\mathrm{v}_d(k_0) \sqrt{M^2 - v^2}}{\xi_{SO}(k_0) v}. \label{eq:lambda_app}
\end{align}
Here, the range of $x$ shall be taken to be a half of the real axis such that $x/\lambda>0$ for $\tilde{\chi}_1$, and $x/\lambda<0$ for $\tilde{\chi}_2$. Changing to the original basis defined by $\bm{d}_{x}$ and $\bar{\bm{d}}_{x}$, we obtain the Majorana zero mode solutions presented in Eq.~\eqref{eq:wf_maj_chain}.

\section{Solutions of chains with finite spin-orbit coupling}\label{app:soc}

In this appendix we deal with the case of non-vanishing spin-orbit coupling in Hamiltonian~\eqref{eq:Hd}. We will limit our analysis to the subgap regime ($|E|<\Delta$) such that the imaginary part of the self-energy in Eq.~\eqref{eq:Gd0} can be dropped. Let us start with the explicit form of the $d$-orbital Green's function by substituting Eq.~\eqref{eq:Hd} into Eq.~\eqref{eq:Gd0},
\begin{align}
  &G_d(E^+) = \left[E^+ + v_E - M\sigma_z + (\mu_d - \xi_{SO}\sigma_y)\tau_z +v_{\Delta}\tau_x\right]^{-1},\label{eq:Gd2}
\end{align}
where $v_E = v E/\sqrt{\Delta^2 - E^2}$, $v_\Delta = v \Delta/\sqrt{\Delta^2 - E^2}$, and we have suppressed the $k_x$ dependence of $G_d$, $\mu_d$ and $\xi_{SO}$ to shorten the expression. The poles of $G_d$ are given by the solutions of
\begin{align}
  &0 = \text{Det}\left[G_d(E)^{-1}\right]
  \simeq (v_E^2 - \epsilon_+^2) (v_E^2 - \epsilon_-^2), \label{eq:D2}
\end{align}
where we have defined the functions
\begin{align}
  &\epsilon_{\pm} = \sqrt{M^2+\xi_{SO}^2+\mu_d^2+v_\Delta^2 \pm 2\sqrt{(M^2+\xi_{SO}^2)\mu_d^2+M^2v_\Delta^2}}, \label{eq:eps_pm}
\end{align}
and we have used the approximation $E+v_E \simeq v_E$ since $|E|< \Delta \ll v$. By solving $v_E^2 = \epsilon_+^2$ and $v_E^2 = \epsilon_-^2$ from Eq.~\eqref{eq:D2}, we obtain 4 solutions. The two solutions of $v_E^2 = \epsilon_+^2$ are outside the gap. Only the two solutions of $v_E^2 = \epsilon_-^2$ satisfy $|E|<\Delta$. Therefore we obtain the spectrum of the Shiba bands to be (with the $k_x$ dependence suppressed)
\begin{align}
  \frac{E_{\pm}}{\Delta} \simeq \pm\sqrt{\frac{[(M^2+\xi_{SO}^2)-(\mu_d^2+v^2)]^2+4\xi_{SO}^2 v^2}{[(M^2+\xi_{SO}^2)-(\mu_d^2+v^2)]^2+4(M^2+\xi_{SO}^2)v^2}}. \label{eq:shiba_band_en}
\end{align}

Next we analyze the topological property of the chain with the assumption that the Shiba bands are fully gapped. This assumption is equivalent to $\forall k_x:$ $E_{\pm}(k_x) \ne 0$, or
\begin{align}
  \forall k_x:\quad Z \equiv M^2+\xi_{SO}^2 - \mu_d^2 - v^2 + 2i\xi_{SO} v \ne 0. \label{eq:Z}
\end{align}
The Green's function in Eq.~\eqref{eq:Gd2} respects two anti-unitary symmetries
\begin{align}
  &\hspace{-0.018\textwidth} T G_d(k_x, E) T^{-1} = G_d(-k_x, E), \quad T=K, \label{eq:T_symm}\\
  &\hspace{-0.018\textwidth} P G_d(k_x, E) P^{-1} = -G_d(-k_x, -E), \quad P=\sigma_y\tau_y K, \label{eq:P_symm}
\end{align}
where $K$ stands for complex conjugation, and we have used the symmetry properties $\xi_d(k_x) = \xi_d(-k_x)$ and $\xi_{SO}(k_x) = -\xi_{SO}(-k_x)$. Physically, $T$ is a combination of mirror symmetry and time-reversal symmetry, and acts like a spinless time-reversal symmetry effectively. These symmetries imply that our model belongs to the BDI symmetry class. It follows that we can bring $G_d(k_x, E)$ at $E=0$ to the following off-block-diagonal form
\begin{align}
  U^\dag\, G_d(k_x, E=0)\, U =
  \begin{pmatrix}
    0 & Q(k_x) \\
    Q(k_x)^\dag & 0
  \end{pmatrix}^{-1}
\end{align}
where
\begin{align}
  &U = \frac{1}{\sqrt{2}}\left[\sigma_0\otimes
  \begin{pmatrix}
    1 & 0 \\
    0 & i \\
  \end{pmatrix}
  +\sigma_y\otimes
  \begin{pmatrix}
    0 & -1 \\
    i & 0 \\
  \end{pmatrix}\right], \\
  &Q(k_x) = [\xi_{SO}(k_x)+iv]\sigma_0 - [iM\sigma_x + \mu_d(k_x)\sigma_y].
\end{align}
Note that the choice of the above transformation is not unique, and we have chosen $U$ such that $Q(k_x)$ becomes purely imaginary at $k_x = 0, \pi$ (where $\xi_{SO}$ vanishes). The topological invariant of our model is thus given by the winding number
\begin{align}
  n &= i\int_{-\pi}^{\pi} \frac{dk_x}{2\pi}\, \mathrm{Tr}\left[Q(k_x)^{-1} \partial_{k_x} Q(k_x)\right] \\
  &=  i\int_{-\pi}^{\pi} \frac{dk_x}{2\pi}\, \frac{\partial_{k_x} Z}{Z},
\end{align}
where $Z$ is given by Eq.~\eqref{eq:Z} and is non-vanishing for all $k_x$ by assumption. This winding number can in principle be an arbitrary integer depending on the specific forms of $\xi_d(k_x)$ and $\xi_{SO}(k_x)$. For generic spin-orbit coupling, however, we assume that $\xi_{SO}$ vanishes only at $k_x=0$ or $\pi$, which implies that $Z$ becomes real only at these momenta. As a consequence, the winding number can only be $0$ or $\pm 1$, and the condition of nontrivial topology assumes a simple form
\begin{align}
  \mathrm{sgn}[Z(k_x=0)\,Z(k_x=\pi)] = -1, \label{eq:topo_cond}
\end{align}
or explicitly, $\mathrm{sgn}[M^2 - \mu_d(0)^2 - v^2]\cdot\mathrm{sgn}[M^2 - \mu_d(\pi)^2 - v^2] = -1$.
Note that this condition is equivalent to
\begin{align}
  \mathrm{sgn}\left[\mathrm{Pf}
  \begin{pmatrix}
    0 & Q(0) \\
    Q(0)^\dag & 0
  \end{pmatrix}
  \mathrm{Pf}
  \begin{pmatrix}
    0 & Q(\pi) \\
    Q(\pi)^\dag & 0
  \end{pmatrix}\right] = -1,
\end{align}
where $\mathrm{Pf}$ stands for Pfaffian. The Pfaffian condition, however, applies even without the $T$ symmetry in Eq.~\eqref{eq:T_symm}. Note also that as long as Eq.~\eqref{eq:topo_cond} is true, there exists at least one pair of momenta $\pm k_0$ ($k_0 \ne 0,\pi$) such that $\mathrm{Re}[Z(\pm k_0)] = M^2+\xi_{SO}(\pm k_0)^2 - \mu_d(\pm k_0)^2 - v^2 = 0$.

In the following we derive the spin densities associated with the Shiba bands and the Majorana zero modes. From Eq.~\eqref{eq:Gd2} we obtain the spectral function
\begin{align}
  A(E) &= \frac{i}{2\pi}[G_d(E^+) - G_d(E^-)] \nn\\
  &\simeq \frac{\psi\psi^\dag}{\psi^\dag\psi}\,[\delta(v_E - \epsilon_-)+\delta(v_E +  \epsilon_-)] \label{eq:A4}
\end{align}
where
\begin{align}
  \psi =
  \begin{pmatrix}
    \xi_{SO}^2\xi_{++} + v_\Delta^2\xi_{--} - \xi_{++}\xi_{--}\xi_{+-} \\
    i\xi_{SO}(\xi_{SO}^2 + v_\Delta^2 - \xi_{--}\xi_{+-}) \\
    -v_\Delta(\xi_{SO}^2 + v_\Delta^2 - \xi_{++}\xi_{+-}) \\
    -i\xi_{SO}v_\Delta(\xi_{++} - \xi_{--})
  \end{pmatrix} \label{eq:psi4}
\end{align}
with $\xi_{\pm\pm} = v_E\pm M\pm\mu_d$, and we have again used the approximation $E+v_E \simeq v_E$. The delta functions in Eq.~\eqref{eq:A4} impose a constraint $v_E^2 = |Z|^2/4M^2$, with $Z$ defined in Eq.~\eqref{eq:Z}, by using the explicit form of $\epsilon_-$ in Eq.~\eqref{eq:eps_pm}. With this constraint, and by using the definition Eq.~\eqref{eq:rho_def} for the $d$-orbital spin densities, we obtain Eqs.~\eqref{eq:rho_subgap_so} and \eqref{eq:rho_A} in the main text. Note that Eqs.~\eqref{eq:A4} and \eqref{eq:psi4} both contain $k_x$-dependence implicitly.

To find the Majorana zero mode solutions, we investigate the spectral function $A$ in Eq.~\eqref{eq:A4} at $E=0$, and extend the domain of $A$ to complex $k_x$ by analytic continuation. We will denote the complex $k_x$ by $\tilde{k}_x$. At $E=0$, we have $v_E = 0$ and $v_\Delta = v$, therefore Eqs.~\eqref{eq:A4} and \eqref{eq:psi4} become
\begin{align}
  &A(E=0) \simeq \frac{2\psi_{0}\psi_{0}^\dag}{\psi_{0}^\dag\psi_{0}}\,\delta\left[\epsilon_-(E=0)\right], \label{eq:A4_E0} \\
  &\psi_{0} \equiv \psi({E=0}) =
  \begin{pmatrix}
    (M+\mu_d)(M^2 + \xi_{SO}^2 - \mu_d^2 - v^2) \\
    i\xi_{SO}(M^2 + \xi_{SO}^2 - \mu_d^2 + v^2) \\
    v(M^2 - \xi_{SO}^2 - \mu_d^2 - v^2) \\
    -2i\xi_{SO}v(M+\mu_d)
  \end{pmatrix}.
\end{align}
From Eq.~\eqref{eq:eps_pm}, the condition $\epsilon_-(E=0)=0$ leads to
\begin{align}
  M^2 + \xi_{SO}(\tilde{k}_x)^2 - \mu_d(\tilde{k}_x)^2 - v^2 = \pm 2i\xi_{SO}(\tilde{k}_x)v, \label{eq:cond_eps=0}
\end{align}
where $\tilde{k}_x$ is complex. This equation is to be contrasted with Eq.~\eqref{eq:Z} where $k_x$ can only be real. Because $\xi_{d}$ (and hence $\mu_d$) is even with respect to $k_x$ and $\xi_{SO}$ is odd with respect to $k_x$, we find the solutions of Eq.~\eqref{eq:cond_eps=0} occur in pairs for each sign on the RHS: if $\tilde{k}_x = \tilde{k}_0$ is a solution of the equation with the plus sign, then $\tilde{k}_x = -\tilde{k}_0^*$ is also a solution for the plus sign, whereas $\tilde{k}_x = -\tilde{k}_0$ and $\tilde{k}_x = \tilde{k}_0^*$ are two solutions of the equation with the minus sign. For simplicity, we will assume there exists and only exists one quadruple of such solutions, denoted by $\pm \tilde{k}_0$ and $\pm \tilde{k}_0^*$ with $\tilde{k}_0 = k_0 + i k'_0$ (both $k_0$ and $k'_0$ are real and finite), to Eq.~\eqref{eq:cond_eps=0}.
It follows that $\psi_{0}$ can be further simplified at this quadruple of complex momenta
\begin{align}
  &\hspace{-0.02\textwidth}
  \psi_{0}(\tilde{k}_x = \tilde{k}_0) = \psi_{0}(\tilde{k}_x = -\tilde{k}_0^*)^* =
  \begin{pmatrix}
    M+\tilde{\mu}_d \\
    v + i\tilde{\xi}_{SO} \\
    v +i\tilde{\xi}_{SO} \\
    -(M+\tilde{\mu}_d)
  \end{pmatrix}, \\
  &\hspace{-0.02\textwidth}
  \psi_{0}(\tilde{k}_x = -\tilde{k}_0) = \psi_{0}(\tilde{k}_x = \tilde{k}_0^*)^* =
  \begin{pmatrix}
    -(M+\tilde{\mu}_d) \\
    v + i\tilde{\xi}_{SO} \\
    -(v +i\tilde{\xi}_{SO}) \\
    -(M+\tilde{\mu}_d)
  \end{pmatrix},
\end{align}
where $\tilde{\mu}_d\equiv\mu_d(\tilde{k}_0)$, $\tilde{\xi}_{SO}\equiv\xi_{SO}(\tilde{k}_0)$, and we have ignored a factor $2i\xi_{SO}v$ because it will always be canceled by the normalization factor ($\psi^\dag\psi$) as in Eq.~\eqref{eq:A4_E0}.  These solutions represent four zero-energy evanescent modes in the chain whose wavefunctions are proportional to $\psi_{0}(\tilde{k}_x)\,\exp(i \tilde{k}_x x)$ with $\tilde{k}_x$ replaced with each of the quadruple momenta.

The solutions of Majorana zero modes can be obtained by superposing these evanescent modes with appropriate open boundary conditions -- in our general model, the boundary condition is not unique.
We find the Majorana zero mode solutions to be (up to a normalization factor)
\begin{subequations}\label{eq:wf_maj_so_supp}
\begin{align}
  \chi_1(x) &=
  \frac{1}{2i}\left[\psi_{0}(\tilde{k}_0) e^{i (\tilde{k}_0 x+\varphi)} - \psi_{0}(-\tilde{k}_0^*)  e^{-i (\tilde{k}_0^* x+\varphi)}\right] \nn \\
  &=
  \begin{pmatrix}
    \chi_{\uparrow}(x) \\
    \chi_{\downarrow}(x) \\
    \chi_{\downarrow}(x) \\
    -\chi_{\uparrow}(x)
  \end{pmatrix} e^{-k'_0 x}, \quad (k'_0 x>0) \\
  \chi_2(x) &=
  \frac{1}{2}\left[\psi_{0}(\tilde{k}_0^*)  e^{i (\tilde{k}_0^* x-\varphi)} - \psi_{0}(-\tilde{k}_0) e^{-i (\tilde{k}_0 x-\varphi)}\right] \nn \\
  &= i
  \begin{pmatrix}
    \chi_{\uparrow}(-x) \\
    -\chi_{\downarrow}(-x) \\
    \chi_{\downarrow}(-x) \\
    \chi_{\uparrow}(-x)
  \end{pmatrix} e^{k'_0 x}, \quad (k'_0 x<0)
\end{align}
\end{subequations}
where $\varphi$ is a phase depending on the boundary condition (assumed to be the same for $\chi_1$ and $\chi_2$),
\begin{subequations}\label{eq:chi_ud}
\begin{align}
  &\hspace{-0.02\textwidth}
  \chi_{\uparrow}(x) = \mathrm{Im}[(M+\tilde{\mu}_d) e^{i (k_0 x+\varphi)}], \\
  &\hspace{-0.02\textwidth}
  \chi_{\downarrow}(x) = \mathrm{Im}[(v+i\tilde{\xi}_{SO}) e^{i (k_0 x+\varphi)}],
\end{align}
\end{subequations}
and $k_0$ and $k'_0$ are the real part and the imaginary part of $\tilde{k}_0$, respectively.

To proceed, we assume $|k'_0|\ll |k_0|$. This assumption is certainly valid in the case of perturbative spin-orbit coupling because generically $k_0 \sim 1/a$, whereas from Eq.~\eqref{eq:lambda_app}, we have $k'_0 = 1/\lambda \sim \frac{\xi_{SO} v}{t_d M}\,(1/a) \ll 1/a$ (here $t_d$ stands for the bandwidth of the pristine $d$-orbit bands and $a$ stands for the lattice constant). With this assumption we can expand $\tilde{\mu}_d=\mu_d(\tilde{k}_0)$ and $\tilde{\xi}_{SO}=\xi_{SO}(\tilde{k}_0)$ to the leading order in $k'_0/k_0$, and we have
\begin{subequations}\label{eq:mud_xiso}
\begin{align}
  &\tilde{\mu}_d \simeq {\mu}_d(k_0) - i\mathrm{v}_d(k_0)k'_0, \\
  &\tilde{\xi}_{SO} \simeq \xi_{SO}(k_0) + i\dot{\xi}_{SO}(k_0) k'_0,
\end{align}
\end{subequations}
where $\mathrm{v}_d(k_0) = (\partial \xi_d/\partial k_x)|_{k_x=k_0} = -(\partial {\mu}_d/\partial k_x)|_{k_x = k_0}$ and $\dot{\xi}_{SO}(k_0) = (\partial {\xi}_{SO}/\partial k_x)|_{k_x = k_0}$. From now on, we shall assume $\xi_{SO}$ is slowly varying around $k_0$ and therefore set $\dot{\xi}_{SO}(k_0) \simeq 0$. Substituting $\tilde{k}_x = k_0 + i k'_0$ into Eq.~\eqref{eq:cond_eps=0} (with the plus sign on the RHS) and expand the equation again to the leading order in $k'_0/k_0$, we further have
\begin{align}
  &M^2 + \xi_{SO}(k_0)^2 - \mu_d(k_0)^2 - v^2 \simeq 0, \label{eq:k0real} \\
  &\mu_d(k_0) \mathrm{v}_d(k_0) k'_0 \simeq \xi_{SO}(k_0)v, \label{eq:k0imag}
\end{align}
where the two equations correspond to the real part and the imaginary part of Eq.~\eqref{eq:cond_eps=0}, respectively. The existence of solutions of Eq.~\eqref{eq:k0real} is ensured by the topological condition Eq.~\eqref{eq:topo_cond}. The combination of Eqs.~\eqref{eq:k0real} and \eqref{eq:k0imag} gives (assuming $\mu_d(k_0)>0$)
\begin{align}
  k'_0 \simeq \frac{\xi_{SO}(k_0)v}{\mathrm{v}_d(k_0)\sqrt{M^2 + \xi_{SO}(k_0)^2 - v^2}}, \label{eq:k0prime}
\end{align}
which reduces to Eq.~\eqref{eq:lambda_app} by neglecting $\xi_{SO}(k_0)^2$ and by noticing that $k'_0$  is equivalent to $1/\lambda$. Furthermore, by substituting Eqs.~\eqref{eq:mud_xiso} [with $\dot{\xi}_{SO}(k_0) \simeq 0$] and \eqref{eq:k0prime} into Eq.~\eqref{eq:chi_ud}, we obtain
\begin{subequations}
\begin{align}
  &\hspace{-0.02\textwidth}
  \chi_{\uparrow}(x) \simeq [M+{\mu}_d(k_0)]\sin (k_0 x+\varphi) \nn\\
  &\qquad\;- [\xi_{SO}(k_0)v/{\mu}_d(k_0)]\cos (k_0 x+\varphi), \\
  &\hspace{-0.02\textwidth}
  \chi_{\downarrow}(x) \simeq v\sin (k_0 x+\varphi) + \xi_{SO}(k_0)\cos (k_0 x+\varphi).
\end{align}
\end{subequations}
It is easy to check that by neglecting the terms proportional to $\xi_{SO}(k_0)$ in the above equations and neglecting $\xi_{SO}(k_0)^2$ in Eq.~\eqref{eq:k0real}, the Majorana zero mode solutions in Eq.~\eqref{eq:wf_maj_so_supp} with $\varphi=0$ become equivalent to those in Eq.~\eqref{eq:wf_maj_chain}. We set $\varphi=0$ in the main text, but we emphasize that the possible dependence of $|\chi_{1}(x)|^2/|\chi_{2}(x)|^2$ on boundary conditions ($\varphi$) is motivation to define the ratio of the integrated Majorana spin densities, which is $\varphi$-independent, in Eq.~\eqref{eq:rhoM_ratio_so}.

\section{Derivation of differential conductances with the set-point effect}\label{app:cond}

The set-point effect is a constraint on the trajectory of the STM tip, such that for each in-plane coordinate $(x,y)$, the height of the tip $z$ is determined by requiring the total tunneling current at a particular bias $V_c = E_c/e$ to be a constant $I_0$. Since we are only interested in the measurements on the chain (at $y=0$), we will ignore dimension-$y$ from now on. By using Eqs.~\eqref{eq:I_chain}, the constraint from the set-point effect reads
\begin{align}
  &I_0 = \sum_{\sigma=\uparrow,\downarrow}w_{N/P,\sigma}(z_{N/P}(x)) R_{\sigma}(x), \\
  &R_{\uparrow/\downarrow}(x) = \int_0^{E_c} dE\; \rho_{\uparrow/\downarrow}(x,E),
\end{align}
where $z_{N/P}(x)$ is the height of the tip with N or P polarization in actual measurements at $x$. With the assumption that the spin-dependence and the $z$-dependence of the weight factors ($w$'s) are separable (see Sec.~\ref{sec:stm} in the main text for a discussion), the above equation can be rewritten as
\begin{align}
  I_0 = w_{N/P,\downarrow}(z_{N/P}(x)) \Bigl[\tilde{w}_{N/P} R_{\uparrow}(x) + R_{\downarrow}(x)\Bigr], \label{eq:I_const2}
\end{align}
where $I_0$ is the constant current, and $\tilde{w}_{N/P} = {w_{N/P,\uparrow}(z)}\bigl/{w_{N/P,\downarrow}(z)}$ is independent on $z$.

Now, by definition the differential conductances
\begin{align}
  &G_{N/P}(x,E) \equiv \left.\frac{\partial I_{N/P}(x,z_{N/P}(x),V)}{\partial V}\right|_{eV=E} \\
  &\;= e\, w_{N/P,\downarrow}(z_{N/P}(x)) \Bigl[\tilde{w}_{N/P} \rho_{\uparrow}(x, E) + \rho_{\downarrow}(x, E)\Bigr] \\
  &\;= (eI_0)\,\frac{\tilde{w}_{N/P}\,\rho_{\uparrow}(x,E)+\rho_{\downarrow}(x,E)}{\tilde{w}_{N/P}\,R_{\uparrow}(x)+R_{\downarrow}(x)},
\end{align}
where we have used Eqs.~\eqref{eq:I_chain} and \eqref{eq:I_const2}, respectively, in the last two steps. After dropping the constant factor $eI_0$, we obtain Eq.~\eqref{eq:Gmodel} in the main text.


%

\end{document}